\documentclass[12pt,a4paper]{iopart}

\usepackage{amstext}
\usepackage{pstricks}
\usepackage{graphicx}
\usepackage{subcaption}
\usepackage{color}
\usepackage{soul}
\usepackage[font=small, labelfont=bf]{caption}
\usepackage{MnSymbol}
\usepackage{subfiles}
\usepackage{cite}
\usepackage{hyperref}
\usepackage{xcolor}
\usepackage{amsmath}
\usepackage{amsfonts}

\usepackage{etoolbox}
\makeatletter
\def\@mkboth#1#2{}
\newlength\appendixwidth
\preto\appendix{\addtocontents{toc}{\protect\patchl@section}}
\newcommand{\patchl@section}{%
  \settowidth{\appendixwidth}{\textbf{Appendix }}%
  \addtolength{\appendixwidth}{1.5em}%
  \patchcmd{\l@section}{1.5em}{\appendixwidth}{}{\ddt}%
}
\makeatother

\newcommand{\dd}{\mathrm{d}}
\newcommand{\avgc}[1]{\langle #1 \rangle_c}
\newcommand{\Ns}{\mathcal{N}_s}
\newcommand{\intR}{\int_{-\infty}^\infty}

\begin{document}

\title{A new spin on optimal portfolios and ecological equilibria}

\author{J\'er\^ome Garnier-Brun$^{1,2}$, Michael Benzaquen$^{1,2,3}$, Stefano Ciliberti$^3$ and Jean-Philippe Bouchaud$^{1,3,4}$.}
\address{$^1$Chair of Econophysics \& Complex Systems, \'Ecole polytechnique, 91128 Palaiseau Cedex, France}
\address{$^2$LadHyX UMR CNRS 7646, \'Ecole polytechnique, 91128 Palaiseau Cedex, France}
\address{$^3$Capital Fund Management, 23 Rue de l’Universit\'e, 75007 Paris, France}
\address{$^4$Académie des Sciences, 23 Quai de Conti, 75006 Paris, France}
\eads{michael.benzaquen@polytechnique.edu}

\date{\today}

\begin{abstract}
We consider the classical problem of optimal portfolio construction with the constraint that no short position is allowed, or equivalently the valid equilibria of multispecies Lotka-Volterra equations with self-regulation in the special case where the interaction matrix is of unit rank, corresponding to species competing for a common resource. We compute the average number of solutions and show that its logarithm grows as $N^\alpha$, where $N$ is the number of assets or species and $\alpha \leq 2/3$ depends on the interaction matrix distribution. We conjecture that the most likely number of solutions is much smaller and related to the typical sparsity $m(N)$ of the solutions, which we compute explicitly. We also find that the solution landscape is similar to that of spin-glasses, i.e. very different configurations are quasi-degenerate. Correspondingly, ``disorder chaos'' is also present in our problem. We discuss the consequence of such a property for portfolio construction and ecologies, and question the meaning of rational decisions when there is a very large number ``satisficing'' solutions.

\end{abstract}



\maketitle

\tableofcontents


\clearpage

\section{\label{sec:intro} Introduction}
When considering a large number of interacting entities, regardless of their nature, statistical physics has proved extremely efficient at gaining both qualitative and quantitative insights on typical behaviour, statistics, and unexpected phase transitions. While many problems can be mapped almost directly to magnetic systems (e.g. neural networks where weights are analogous to the mean magnetisation on discrete lattice sites \cite{amit1985spin,gardner1987zero,Hwang2019}), other systems are constrained by conditions that are less natural in physical systems. An example of such constraint is that of the non-negativity of some quantity, like in the famous perceptron model and the associated storage problem \cite{rosenblatt1958perceptron,engel2001statistical}. In this paper, we address  two further examples of systems with this restriction: (i) ``long-only'' optimal portfolios of volatile financial assets, and (ii) equilibrium populations of competing species. 

The idea of maximising gains while minimising risk for a portfolio of fluctuating assets is one that is both at the origin and at the heart of quantitative finance. As early as 1952, Harry Markowitz derived the general formula for the portfolio with the smallest variance for a desired return \cite{Markowitz1952}, assuming the asset fluctuations are fully described by a covariance matrix. Since then, Markowitz portfolio theory has stood as a reference in portfolio management, and questions surrounding  optimal portfolios have been a very successful playground for physicists. By adapting calculations from the physics of disordered systems, several theoretical results have been obtained, mainly around the phase transition observed when the time series used to infer the covariance matrix become too short relative to the size of the portfolio \cite{ciliberti2007risk,Varga-Haszonits2016,Kondor2017}, and the impact of having noisy covariance matrices in general \cite{Pafka2002,Pafka2004, bun2017cleaning}. The effect of having further constraints in the optimisation problem has also been explored, with very rich results. In particular, imposing that investors must provide a deposit proportional to the value of the underlying assets was shown to result in an exponential number of valid locally optimal portfolios, reminiscent of the number of metastable states in a spin-glass \cite{Galluccio1998}. As argued in that paper, the existence of a very large number of nearly degenerate (or ``satisficing'') solutions for a given optimisation problem is conceptually important because common knowledge can no longer be assumed. The complexity of the problems faced by the agents generates irreducible uncertainty, a quandary called ``radical complexity'' by one of us.\footnote{see, e.g.  \url{www.res.org.uk/resources-page/radical-complexity.html} .}

A constraint that has long attracted much interest in the risk management industry consists in enforcing \textit{long-only} portfolios. Denoting $\mathbf{w} \in \mathbb{R}^N$ the vector of weights associated to each of the $N$ possible assets an investor can consider with $\sum_i w_i = 1$ (fully invested),  a \textit{long} position corresponds to a weight $w_i > 0$, and conversely a \textit{short} sell refers to $w_i < 0$. There are a variety of reasons why one might need to avoid short sells, ranging from explicit investment mandates to extreme cases such as regulatory bans as those seen in Europe during the Coronavirus outbreak. On the fundamental level, very interesting behaviour has  been observed when the long-only constraint is enforced, as portfolios quickly tend to become very sparse, resembling choices made by individual stock pickers \cite{Clarke2011,Lehalle2021portfolio,reigneron2020agnostic}.

In the context of population dynamics, the non-negativity constraint is quite intuitive, as the number of individuals in a given species can obviously not be smaller than zero. Interestingly however, this quantity does not result from a constrained optimisation procedure as in the portfolio case, but rather from a dynamical process governed by ordinary differential equations. Despite this difference, statistical physics-inspired calculations have also resulted in key insights~\cite{PhysRevLett.118.048103,advani2018statistical,Landmann2018,biroli2018marginally,roy2019numerical}. Recent work around the generalised Lotka-Volterra model, representing interacting species evolving in a finite environment, demonstrated the existence of different phases of either unbounded growth, multiple unstable attractors, or a unique stable equilibrium \cite{bunin2017ecological}, depending on the shape of interactions. To exploit the formal similarity with long-only optimal portfolios, we shall focus on the last phase where a stable equilibrium can be reached, leaving the dynamical picture aside. 

The objective of this paper is to study how the width and shape of the disorder distribution -- here asset volatility or species interactions -- affects the features of the long-only optimal portfolio and the equilibrium populations respectively.\footnote{While seemingly similar to our work due to the long-only constraint, Ref. \cite{Kondor2017} considers uncorrelated assets described by empirical correlation matrices, which makes the two studies very different from one another.} Two quantities that will particularly interest us are (i) the fraction of elements in the solution that are nonzero (representing the sparsity of the portfolio or the fraction of surviving species), and (ii) the number of acceptable solutions satisfying the constraint (degeneracy).
Our results contrast the standard (spin-glass like) optimisation problem in two ways. First, the {\it average} number of solutions grows sub-exponentially with $N$, in a way that depends sensitively on the nature of the disorder. Second, the {\it typical} (i.e. most likely) number of solutions is very different from (and much smaller than) the average.  

In Section~\ref{sec:portfolio} we introduce the underlying models for these two seemingly  different problems. Due to the availability of high quality empirical data for the portfolio problem, Sections~\ref{sec:numerical} to \ref{sec:results}  focus on disorder distributions compatible with financial assets, and present numerical and  analytical descriptions of the quantities of interest. Distribution-specific results for normally and uniformly distributed disorder are  extended to a generalised normal distribution.  Section~\ref{sec:chaos} numerically establishes disorder chaos, which is interpreted in the context of both asset management and ecological equilibria. In Section~\ref{sec:conclusion}, we conclude and discuss future directions.

\section{\label{sec:portfolio}Underlying models}

\subsection{Portfolio optimisation}

Consider a portfolio of $N$ single assets, and assume (in line with the majority of studies on portfolio optimisation) that asset returns are correlated Gaussian variables.  The portfolio statistics is fully caracterised by its covariance matrix $C_{ij} = \overline{\eta_i \eta_j} - \mu_i \mu_j$, where we have introduced for each asset $i$ the expected return $\mu_i$ and the  fluctuation $\eta_i$. The overline indicates a time average. The full correlation matrix is notoriously difficult to infer from noisy financial time series (see e.g. \cite{bun2017cleaning}), which is why  simplifying hypotheses are generally used in the asset management literature. One of them is given by the one-factor risk model, which, rather than attempting to incorporate all possible sources of fluctuations, assumes that correlations are mostly due to the market exposure. In this framework the covariance matrix writes:
\begin{equation}
    C_{ij} = z_i \delta_{ij} + \beta_i \beta_j ,
\end{equation}
where $z_i$ denotes the variance of asset $i$, and $\beta_i$ its correlation to the market, or more precisely the ratio of the return covariance with the market index to the return variance. While this approximation may appear very coarse, empirical analyses on stocks show that the top eigenvalue of the correlation matrix, corresponding to the market mode, is indeed largely dominant relative to the other eigenvalues (see e.g. \cite{laloux1999noise}).

For an investor interested in constructing an optimal portfolio, the expected returns are of course key parameters. However for this theoretical analysis, which aims at drawing qualitative insights regarding the multiplicity of solutions and its implication on portfolio stability, we impose the simplification $\boldsymbol{\mu} = \boldsymbol{1}$, where $\boldsymbol{\mu} =\{\mu_i\}_{i\in[1,N]}$ (see  \ref{appendix:heterogeneous} for extensions to arbitrary $\mu_i$'s). In this case, the optimal portfolio, also coined the Markowitz portfolio, which minimises the risk  $\sigma_p^2 = \sum_{i,j} C_{ij} w_i w_j$ for the given expected return $\mu_p= 1$, can easily be shown to write: 
\begin{equation}
    w_i = \frac{\sum_j C^{-1}_{ij}}{\sum_{i,j}C^{-1}_{ij}}.
    \label{equ:Markowitz}
\end{equation}
Within the one factor model, the covariance matrix can be easily inverted using the Woodbury matrix identity~\cite{hager1989updating}. Up to a normalising constant, one obtains:
\begin{equation}
    w_i \propto \frac{1}{z_i} \left(1 - \beta_i \frac{\sum_{j} \beta_j/z_j}{1 + \sum_{j} \beta_j^2/z_j} \right).
    \label{equ:w_i}
\end{equation}
This  equation is central for the problem that we aim to explore in the following.

\subsection{Ecological equilibria}

Consider now $i=1, \dots,N$ species associated to a certain ``{carrying capacity}'' in the environment. These species furthermore interact with each other, either competing for resources, or in predator-prey relationships, or else in a mutualistic, cooperative mode. 

In its simplest form, where we consider that all species have identical growth rates $\mu_i=1$, the population dynamics are described by the general Lotka-Volterra equations:
\begin{equation}
\partial_t{S}_i(t) = S_i(t) \left[\mu_i - \mu_i k_i^{-1} S_i(t) - \sum_{j} \alpha_{ij} S_j(t) \right],\quad \text{with} \quad \mu_i=1  \quad \forall i,
\label{equ:LV}
\end{equation}
where $S_i$ is the population of species $i$, $k_i$ its carrying capacity, and $\alpha_{ij}$ is the $N \times N$ interaction matrix \cite{bunin2017ecological}.  In this model, a positive entry $\alpha_{ij}$ corresponds to species $i$ and $j$ competing for resources or $i$ being a prey and $j$ being a predator (in which case $\alpha_{ji} < 0$). Setting $\partial_t{S}_i = 0$ to identify fixed points of the system yields the equilibrium population of the species:
\begin{equation}
    S_i = \sum_j C^{-1}_{ij},
    \label{equ:species_equilib}
\end{equation}
where here $C_{ij} = z_i \delta_{ij} + \alpha_{ij}$, with $z_i = k_i^{-1}$. Naturally, in this  context, one must have $S_i \geq 0 \; \forall \, i$  since populations cannot be negative.

Experimentally, it is very difficult to gain insight on the nature of the interaction matrix or its eigenvalues. As a matter of fact, it is this observation that initially motivated Robert May to use Random Matrix Theory arguments in his seminal paper \cite{may1972will}. While the qualitative phase portrait for the dynamical behaviour of the model is independent of the exact distribution of $\alpha_{ij}$ \cite{bunin2017ecological}, it seems natural from Eq.~\eqref{equ:species_equilib} that the equilibrium picture would be dependent on the interaction matrix model.

Here, we propose a drastic simplification and choose the interaction matrix to be of unit rank: $\alpha_{ij} = \beta_i \beta_j$, corresponding to species competing for a single common resource, in addition to the self-regulation included in Eq.~\eqref{equ:LV}. We will take $\beta$'s to be independent and identically distributed, with a probability density function $\rho(\beta)$. The $\beta_i$ coefficients then quantify how strongly species $i$ competes for the unique resource with other species, and the interaction between two species then only depends on how strongly they both depend on the resource. With this model of interactions, the equilibrium populations map to the long-only optimal portfolio weights (up to a constant that does not affect the sign), and both problems can be treated identically based on Eq.~\eqref{equ:w_i}. Note that the case of heterogeneous growth rates $\mu_i$ is equivalent to different average returns for stocks, and is discussed later in \ref{appendix:heterogeneous}.

Taking $\langle \beta \rangle > 0$ is a natural choice, both to avoid placing ourselves in an unbounded growth regime, and more generally because ecosystems tend to be highly competitive. Naturally, $\sigma^2=\mathbb{V}(\beta)$ shall also play a key role in the equilibrium picture of the system and its properties. Finally, it is important to note that the unit rank model yields a symmetric interaction matrix, which amounts to either cooperation or pure resource competition.  Predator-prey relations require, as noted above, asymmetric interactions between species $i$ and $j$.

\subsection{Spin-glasses}

Suppose we now enforce the non-negativity constraint common to the two problems. For the portfolio, this means the positions associated to short sells after the Markowitz optimisation, i.e. associated to $w_i < 0$, will have to be removed from the portfolio altogether, reducing the effective universe from which stocks may be picked. Likewise, an extinct species ($S_i < 0$) is by definition removed from the ecological universe, leading to an ecology with a reduced number of viable species.

Here, we introduce `spins' $\{ \theta \}$ that can take the value $\theta_i = 1$ if position $i$ is included in the (possibly reduced) asset or species universe and $\theta_i = 0$ if it is excluded from it. Clearly, without excluding any specific entity, $2^N$ combinations of $\{ \theta \}$ can be constructed from the $N$ assets or species initially considered\footnote{$2^N - 1$ solutions to be exact, as the empty portfolio cannot satisfy the fully invested constraint.}. The central question treated in this paper can therefore be reformulated as follows: we seek the number $ \mathcal{N}_s $ of  possible configurations $\{\theta\}$, among these $2^N$,  that  satisfy the non-negativity condition. These spin variables are then related to the weights of the underlying positions through Eq.~\eqref{equ:w_i}. Indeed, only the included positions, i.e. those with $\theta_i = 1$, now contribute to the sums, while the weights associate to positions with $\theta_i = 0$ are, by definition, discarded.

This quantity can easily be understood for financial assets, as it corresponds to the number of long-only Markowitz-optimal portfolios that can be constructed from a set of $N$ assets. In the context of ecological equilibria, the interpretation is similar even if species are not ``selected'' in the same way as stocks. $ \Ns $ can then be seen as the number of viable stable ecosystems that result from particular subsets of the $N$ interacting species. The existence of solutions with a lower number of highly concentrated species in addition to the default, most diverse, solution is actually particularly interesting in the ecology context when considering the so-called \textit{Allee effect}, which states that an increase in population density is correlated with higher survival probability \cite{courchamp1999inverse}. In both cases, this quantity, which may appear somewhat artificial at this stage, will be essential in understanding how disorder chaos arises and can impact these systems in a very concrete way.

Naively, one could try to characterise the number of solutions by its average, 
\begin{equation}
    \langle \mathcal{N}_s \rangle = \Big\langle \sum_{\{ \theta \}} \prod_{k=1}^N \Theta \left( \theta_k w_k \right) \Big\rangle,
    \label{equ:direct}
\end{equation}
where we take the convention $\Theta(0) = 1$ for the Heaviside step function, and averages are taken over the distribution of $\beta$. At this stage, readers familiar with the physics of disordered systems may notice how formally similar this enumeration is to counting metastable states in quenched spin-glasses, where the Heaviside step function would be replaced by a Dirac $\delta$ distribution with an argument minimising the Hamiltonian \cite{Castellani2005,Bray1980,Bray1981}. As in such physical systems, a key quantity in the study of the number of solutions is the ``annealed'' complexity
\begin{equation}
    \Sigma = \frac{\log \, \langle \mathcal{N}_s \rangle}{N}.
\end{equation}
In the case of Sherrington-Kirkpatrick spin-glasses, this quantity is indeed equivalent to its more representative ``quenched'' counterpart, where the logarithm is averaged (requiring a much more challenging replica calculation), for metastable states of sufficiently high energy \cite{Bray1980}. Another useful observable is the sparsity, describing the average fraction of the $N$ possible entities that are included in the configuration. We can write it as
\begin{equation}
    u = \frac{1}{N} \sum_{i = 1}^N  \theta_i.
\end{equation}

Among the $\mathcal{N}_s$ configurations satisfying the non-negativity constraint for a given set $\{\beta_i\}$, one will be ``as full as it can be'', meaning $u$ will reach its maximum value, the average value of which is noted $m$ below, with $m \leq 1$. In asset management terms, this quantity corresponds to the most diversified long-only portfolios. 
For population dynamics, it is the most diverse ecosystem that can result from all possible viable ecosystems resulting from the $N$ species.

\section{\label{sec:numerical}Numerical experiments}

\subsection{Empirical data}

In order to study the likely number of valid solutions, it is essential to have some information on the distribution of the $\beta$ coefficients.

\begin{figure}
    \centering
    \includegraphics[width=0.66\linewidth]{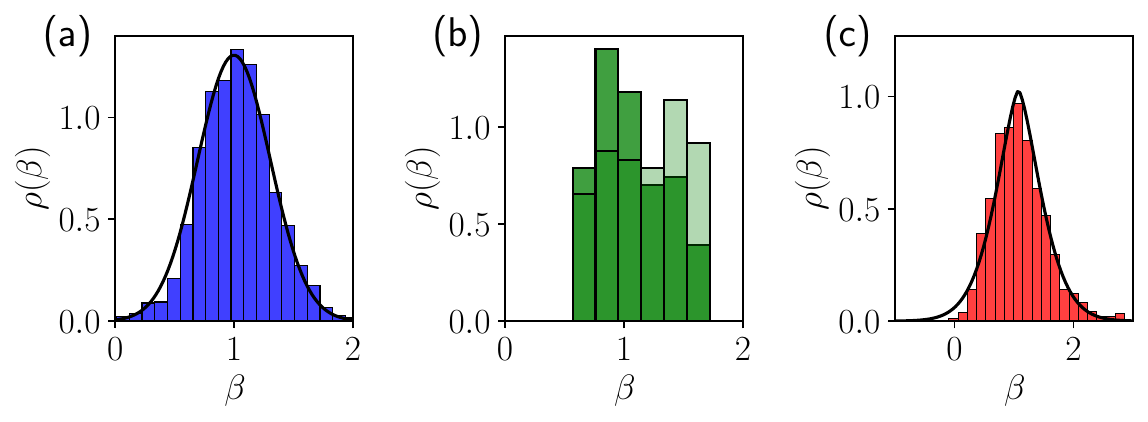}
    \caption{Distributions of $\beta$ computed using the variance and covariance of returns averaged over one and two years respectively (as accurately estimating covariance requires more data). (a) Daily returns of the 1500 largest capitalisation stock, and fit to a normal distribution. (b) (dark green) Daily returns of the 120 largest capitalisation stocks in the energy and utilities industries, (light green) uniformly distributed points over the same interval for the same sample size. (c) Weekly returns of the 1000 largest capitalisation stocks, and fit to a generalised normal distribution of shape parameter $b\approx 3/2$.}
    \label{fig:real_distribs}
\end{figure}

For the portfolio problem, where $\beta$ is a widespread metric for an asset's correlation relative to the market, high quality data is readily available. Using returns from a large number of US stocks over a two year time span (up to November 2020) reveals that the $\beta$s are normally distributed about 1, as shown in Fig.~\ref{fig:real_distribs}(a). We shall thus take as a starting point i.i.d. variables $\beta_i \sim \mathcal{N}(1,\sigma^2)$, which conveniently implies that $\sigma$ and $N$ are the problem's sole parameters.
Nevertheless, all calculations can easily be generalised to any $\langle \beta \rangle \neq 1$ since the problem is invariant under the simultaneous scaling of all $\beta$s and all $z$s by an arbitrary factor $\alpha$ and $\alpha^2$, respectively. 

This being said, other distributions for $\beta$ can also be of interest. In particular, if one focuses on specific   sectors, the empirical distribution of volatility matches a uniform distribution relatively well as can be seen in Figure
\ref{fig:real_distribs}(b). Alternatively, looking at weekly returns rather than daily returns to construct the volatility gives thicker tails, between a Gaussian and Laplace distribution, as well as some slight skewness, visible in Fig.~\ref{fig:real_distribs}(c). 

As mentioned in the previous section, there is unfortunately no such data for the interaction matrix in ecological communities, so in this context our model parameters should be considered with a grain of salt.

In the following numerical experiments, markers labelled by `Data' will be referring to calculations that are using the empirical distributions of $\beta$ from Fig.~\ref{fig:real_distribs}. Practically, the shape and width of the empirical distributions are obtained by fitting the histograms, and data points are then constructed by random sampling with replacement for different values of $N$.

\subsection{Maximum sparsity}

Calculating  the average maximum sparsity~$m$ numerically is rather straightforward with homogeneous returns, as it simply requires to remove iteratively entries for which Eq. \eqref{equ:w_i} gives negative weights, i.e. setting $\theta_i = 0$ for these positions, until all positions are acceptable (see \cite{sankaran1999optimal}).
The result for stock-compatible $\beta \sim \mathcal{N}(1,\sigma^2)$ in Fig.~\ref{Fig:Gaussian}(a) is consistent with the findings of Lehalle and Simon \cite{Lehalle2021portfolio}: the sparsity decreases rapidly and non-trivially when the $\beta$s cease to be very tightly distributed about unity. Interestingly, the sparsity clearly appears to be a function of the parameter $\chi = \sigma N$ only (see Fig.~\ref{Fig:Gaussian}(a)). Such a scaling provides precious insight for the analytical formulation of the problem, as will become apparent in the following section. It should be noted that such a scaling result ceases to hold for large values of $\sigma$, as a large standard deviation yields a significant fraction of assets with negative $\beta$ that can obviously be included in the long-only portfolios (see Eq. \eqref{equ:w_i}), thereby causing $m$ to increase again at large $\sigma$. 

This effect can be partially observed in the empirical points that have a slightly wider distribution of $\beta$ ($\sigma \approx 0.3$) and a few negative entries (not shown). 
This being said, in a range for $N$ and $\sigma$ relevant for applications, the evolution of $m$ for real data points appears to be roughly in line with the $\chi = \sigma N$ scaling curve where the fully numerical points lie. 

The same procedure may be repeated for uniformly distributed $\beta$ as show in Fig.~\ref{Fig:uniform}(a). The result is qualitatively very similar, albeit with a slower decrease in $m$ with $N$. Interestingly we recover the scaling $\chi = \sigma N$ where $\sigma$ now governs the width of the distribution. Like in the Gaussian case, the points  sampled from empirical data appear to be slightly too widely distributed to perfectly match the points from continuous probability densities, although the evolution of $m$ appears to be very close up to some offset.

\subsection{Exact enumeration of solutions}

The other numerical experiment that can be carried out to guide our study of the multiplicity of solutions is an exact enumeration. The procedure relies on testing all $2^N$ possible combinations of $\{ \theta \}$, calculating the weight for each, and counting those that yield only positive weights. Given the exponential number of configurations to be tested, we are limited to $N \leq 32$ in order to keep the computation times reasonable. 

\begin{figure}
    \centering
    \includegraphics[width=\linewidth]{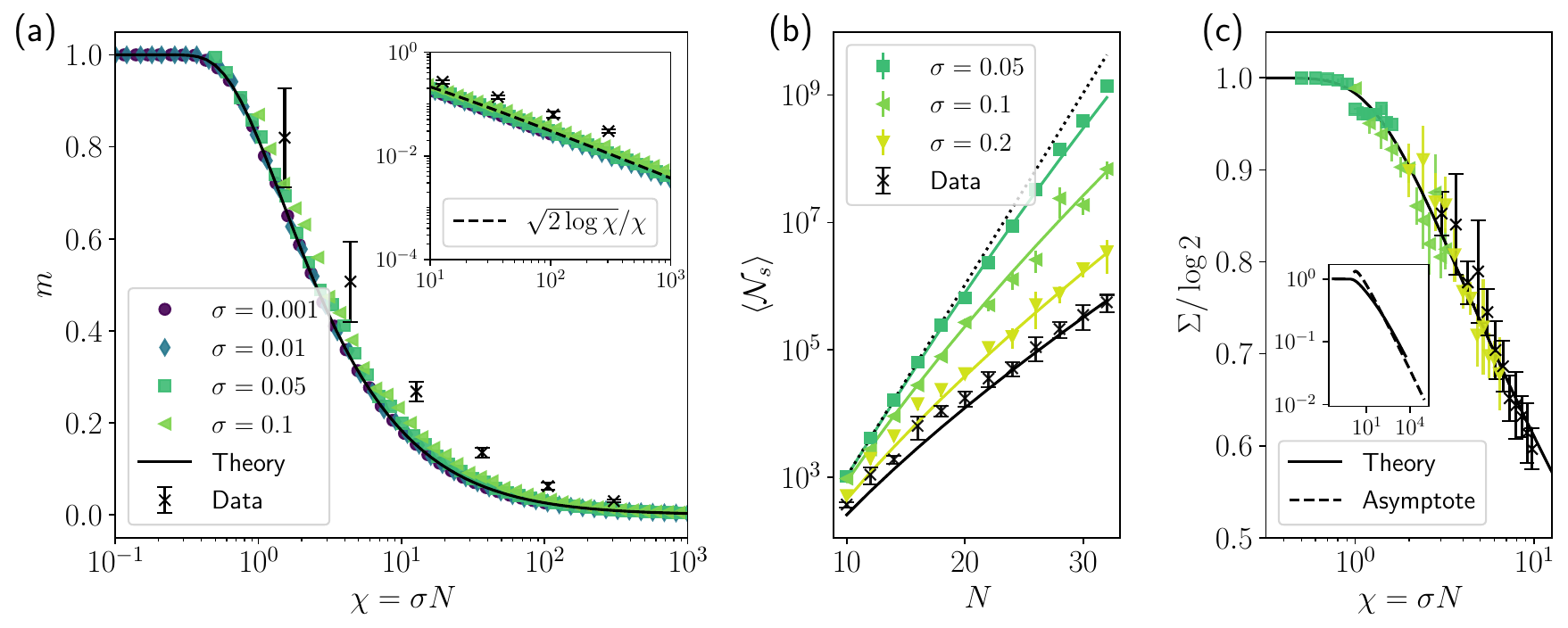}
    \caption{Numerical and theoretical results for $\beta \sim \mathcal{N}(1,\sigma^2)$. (a) Maximum sparsity $m$ as a function of $\chi = \sigma N$, inset focusing on the large $\chi$ region. (b) Number of solutions $\langle \mathcal{N}_s \rangle$ as a function of $N$, obtained by exact enumerations averaged over relatively few (4-10) samples. Straight lines display the respective theoretical predictions, dotted line corresponds to $2^N$. (c) Complexity $\Sigma$ as a function of $\sigma N$ resulting from the exact enumerations and normalised by $\log 2$, inset zooming out to show the large $\chi$ region. The full line is the numerically exact result, and the dotted line is an asymptotic approximation based on Eq. \eqref{eq:advection} below. The legends are shared for (b) and (c).}
    \label{Fig:Gaussian}
\end{figure}

From Fig.~\ref{Fig:Gaussian}(b), we find that, as expected, the number of solutions is close to~$2^N$ when $\sigma \to 0$, and then decreases as the number of negative entries to be removed  increases. The complexity $\Sigma$ for the same datapoints in Fig.~\ref{Fig:Gaussian}(c) gives further insights on the evolution and transition from full to sparse portfolio. Clearly, the complexity also scales almost perfectly with $\chi = \sigma N$.

\begin{figure*}
    \centering
    \includegraphics[width=\linewidth]{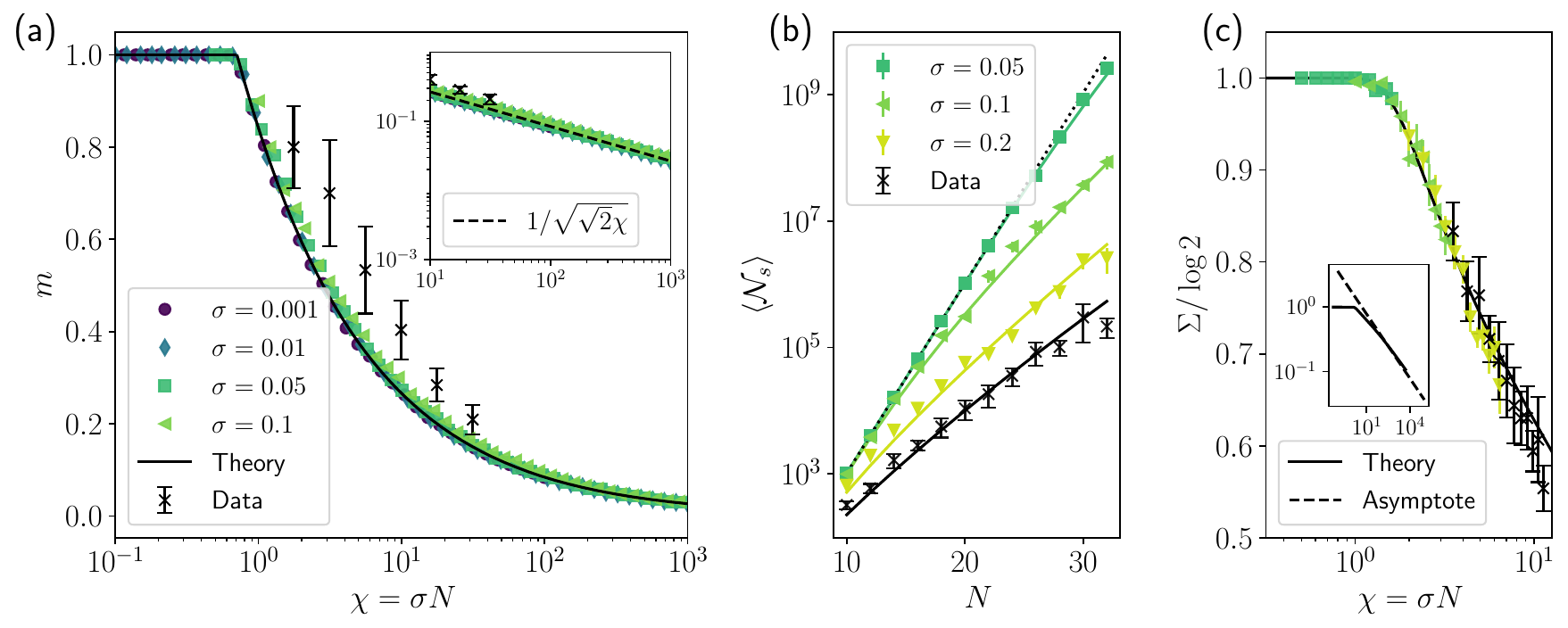}
    \caption{Numerical and theoretical results for $\beta \sim \mathcal{U}(1 \pm \sqrt{2}\sigma)$. See Fig.~\ref{Fig:Gaussian} for the detailed caption.}
    \label{Fig:uniform}
\end{figure*}

Repeating the experiment for the uniform distribution presented in Figs.~\ref{Fig:uniform}(b-c) displays a comparable result, with a slightly slower decrease of $\Sigma$, a difference that is consistent with the previously observed maximum sparsities. 

\subsection{Exploration of the solution space}
\label{subsec:solution_space}
Beyond the macroscopic observables that are the number of valid solutions and the maximum sparsity, the precise distributions of more specific quantities among the long-only configurations may also be explored numerically. Here, two of these are particularly interesting in the context of the multiplicity of solutions.

First, the quadratic form
\begin{equation}
    \mathcal{H}(\{ \beta \},\{ \theta \}) = \sum_{i,j = 1}^N C_{ij} \theta_i w_i \theta_j w_j,
\end{equation}
with the weights $w_i$ solutions to Eq.~\eqref{equ:w_i} with summations restricted to positions with ${\theta_i=1}$, represents the metric that is minimised under  constraint. For the portfolio problem, this is obviously the total portfolio square volatility. Although less straightforward, $\mathcal{H}$ can also be understood intuitively in the population dynamics context. Indeed, as $C_{ij}$ represents the level of competition for the resource between species $i$ and $j$, the quantity to be minimised corresponds to the aggregated level of competition (including self-competition)  for the surviving species. 

The second quantity of interest is the overlap between two configurations $\{\theta \}$ and $\{ \tilde{\theta} \}$. We choose to define it as
\begin{equation} \label{eq:overlap}
    q(\{\theta\},\{\tilde{\theta}\}) = \frac{1}{N}\sum_{i=1}^N \theta_i \tilde{\theta}_i - \left( \frac{1}{N} \sum_{i=1}^N \theta_i \right) \left( \frac{1}{N} \sum_{i=1}^N \tilde{\theta}_i \right).
\end{equation}
This definition differs from the usual spin-glass expression with the addition of the second term on the RHS, that is included to ensure that two statistically independent configurations have zero overlap on average. 

Using these two metrics, we can study the distribution among valid configurations of their excess variance or level of competition relative to the value for the globally optimal configuration $\{ \theta^\star \}$
\begin{equation}
    \Delta \mathcal{H} = \mathcal{H}(\{\beta\},\{ \theta \}) - \mathcal{H}(\{\beta\},\{ \theta^\star \})
\end{equation}
as well as their normalised overlap with this global optimum
\begin{equation}
    \mathcal{Q} = \frac{q(\{\theta\},\{\theta^\star\})}{q(\{\theta^\star\},\{\theta^\star\})}.
\end{equation}
A value of $\mathcal{Q}$ close to 1  indicates a configuration has a large number of common species or assets with the optimal configuration, $\mathcal{Q}$ close to 0 corresponds to solutions much sparser than the optimum while negative values of the overlap are reached for configurations that are largely full but orthogonal to the best possible outcome. 

\begin{figure*}
    \centering
    \includegraphics[width=0.67\linewidth]{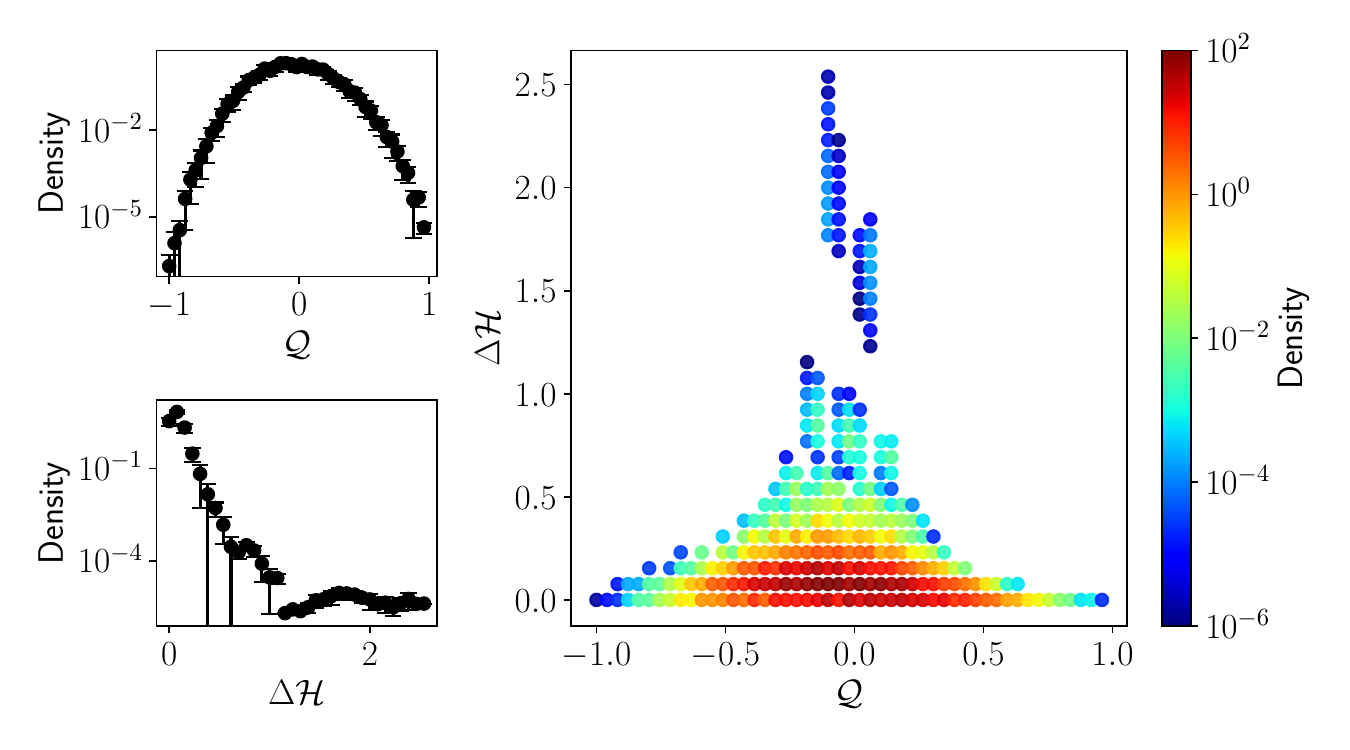}
    \caption{Solution space explored by exact enumeration for Gaussian $\beta$ with $N = 28$, $\sigma = 0.1$, averaged over 30 realisations. Top left: distribution of overlap relative to the optimal configuration $\mathcal{Q}$. Bottom left: distribution of excess variance relative to the optimal configuration $\Delta \mathcal{H}$. Right: heatmap of the joint density of these two quantities.}
    \label{Fig:sol_space}
\end{figure*}

The distributions of $\Delta \mathcal{H}$ and $\mathcal{Q}$ can be obtained directly from the exact enumerations for small values of $N$. Such a result for normally distributed $\beta$ is shown in Fig.~\ref{Fig:sol_space}, which also displays the joint density of the two quantities. Looking at the overlap, we find that, as one could have expected, the majority of valid solutions are composed of a relatively small number of non-zero spins and therefore have $\mathcal{Q}$ close to zero. More surprisingly, looking at $\Delta\mathcal{H}$ reveals that a very large fraction of these are associated to a small excess variance or level of competition relative to the minimum. This is further confirmed by the joint distribution, where we indeed observe that many configurations display a small value of $\Delta \mathcal{H}$ despite having $\mathcal{Q} \approx 0$. As such, the multiplicity of solutions and associated complexity is of a great importance in this problem. Not only do we find a large number of portfolios or ecosystems that satisfy the constraint, a large fraction of these achieve a portfolio variance or level of competition very close to the best possible outcome. These quasi-degenerate solutions might therefore become optimal following a small change in the disorder. This idea is at the root of the disorder chaos investigated in Section \ref{sec:chaos}, and related to the {\it de facto} limitation of rational choice arguments in complex situations.

Note that, interestingly, we can also recover a subset of the solution space displayed in Fig.~\ref{Fig:sol_space} for relatively large values of $N$ by modifying the iterative procedure yielding the optimal configuration (not shown here). To achieve this, we introduce some stochasticity in the algorithm by excluding positions associated to negative weights with a probability $p < 1$ and those associated to positive weights with probability $1-p$. As such, one eventually obtains a non-negative solution, but has randomly removed some spins that could have belonged to the optimal configuration. Repeating this computationally inexpensive procedure a large number of times for fixed disorder, one can find the distribution of $\Delta \mathcal{H}$ for some region in the distribution of $\mathcal{Q}$, that depends  on the value of $p$ chosen. Clearly, taking $p$ close to 1 would only uncover configurations with a normalised overlap close to 1 whereas taking smaller values of $p$ would eventually allow one to explore $\mathcal{Q}$ close to or below zero. While not as detailed and complete as the picture given by the exact enumeration, this method allows us to verify that the key characteristics of the solution space, namely having a large density of solutions in the small $Q$ and $\Delta \mathcal{H}$ region, remain qualitatively similar as $N$ increases significantly. 

As $\Delta \mathcal{H}$ is analogous to the excess energy relative to the ground state in spin-glasses, more complex numerical techniques that have proved effective with SK or EA Hamiltonians (e.g. \cite{schnabel2018distribution}) could also be implemented for a more thorough exploration of the solution space.

\section{\label{sec:self-cons}Analytical setup}
\subsection{Self-consistent equation}

While it may appear natural to attempt to directly tackle the enumeration of valid solutions using Eq.~\eqref{equ:direct} and a Fourier representation of the Heaviside function, such a calculation quickly requires a Gaussian assumption on the distribution of $\beta$ and thus lacks generality.
The alternative taken here is to first study directly the maximum sparsity $m$ as a function of $N$, before translating this quantity back to the number of long-only portfolios.

Going back to Eq.~\eqref{equ:w_i} that relates the weight of position $i$ to its $\beta_i$, it immediately appears that there should be a threshold value $\beta^+$ above which a position will likely be shorted (or a specie go extinct), which must thus be excluded. This is quite reasonable intuitively: an investor wishing to take the least possible risk and unable to balance volatility through shorts will be unlikely to pick excessively risky stocks or bonds. Likewise, since large values of $\beta$ are associated to species subject to increased competition, those species are likely to go extinct and thus not be present in the equilibrium population. Given the distribution $\rho(\beta)$, the average maximum sparsity $m$ is then related to the threshold $\beta^+$ through:
\begin{equation}
    m = \int_{-\infty}^{\beta^+} \mathrm{d}\beta \, \rho(\beta).
    \label{equ:continuous_def}
\end{equation}
Therefore, calculating $\beta^+$ will directly yield $m$.
Of course, this threshold is only valid in a statistical sense, and for a given set of $\beta$'s its value will differ from the mean. Writing $\tilde{\beta}^+$ the fluctuating variable representing the largest $\beta$ to be included for a unique realisation of the disorder, and modifying Eq.~\eqref{equ:w_i}  with the previously introduced spin notation directly gives
\begin{equation}
    \tilde{\beta}^+ = \frac{\sum_j \beta_j^2 \theta_j / z_j + 1}{\sum_j \beta_j \theta_j /z_j},
    \label{equ:beta_init}
\end{equation}
where we now have $\theta_i = 1$ for $\beta_i < \tilde{\beta}^+$ and $\theta_i = 0$ otherwise. In order to recover the typical behaviour of interest, we introduce the probability of inclusion conditioned to $\beta$. We define $\mathrm{Prob}(\theta_i = 1 | \beta_i) = F_N(\beta_i)$, where $F_N(\beta)$ is a smooth step-like function, centred about the mean $\beta^+$. Clearly, we require $F_N(\beta)$ to be monotonously decreasing, with
\begin{equation}
    F_N(-\infty) = 1 \quad \text{and} \quad F_N(\infty) = 0,
\end{equation}
and $F_N'(\beta)$ is therefore peaked in a region around $\beta^+$, the width of which is expected to decrease as $N$ increases.


At this stage, we start by assuming that two spins are uncorrelated at order $N^{-1}$, which shall be checked \textit{a posteriori} by ensuring the Onsager ``reaction'' term is $o(N^{-1})$. The sums at the numerator and denominator of Eq.~\eqref{equ:beta_init} may therefore be treated using the central limit theorem, i.e. for $N\gg 1$
\begin{equation}
    \frac{1}{N} \sum_j \frac{\beta^k_j \theta_j}{z_j} \approx \langle z^{-1} \rangle\llangle \beta^k \rrangle + \frac{1}{\sqrt{N}} \xi_k
\end{equation}
where the partial expectation operator $\llangle \cdots \rrangle$ is defined, for an arbitrary test function $g(.)$, as
\begin{equation}
\llangle g(\beta) \rrangle := \intR \dd \beta \, g(\beta) \rho(\beta) F_N(\beta)
\end{equation}
and $\xi_k$ is a zero-mean Gaussian variable, with a variance that depends on $k$. Eq.~\eqref{equ:beta_init} may then be rewritten as
\begin{equation}
    \tilde{\beta}^+ = \beta^+ + \frac{1}{\sqrt{N}} \xi,
\end{equation}
where the full expressions of $\beta^+$ and $\xi$ are given in \ref{appendix:full_equ}. From the very definition of $F_N(\beta)$, which represents the probability of inclusion in the reduced universe of an asset or specie with respectively correlation or interaction strength $\beta$, one can write $F_N(\beta) = \mathrm{Prob}\left( \beta < \tilde{\beta}^+ \right)$. Using that $\xi$ is a Gaussian noise, this can be expressed as
\begin{equation}
    F_N(\beta) = \frac{1}{2} \operatorname{erfc}\left[ \frac{\sqrt{N}(\beta-\beta^+)}{\gamma \sqrt{2}} \right],
    \label{equ:F_N}
\end{equation}
where $\gamma$ is the standard deviation of $\xi$.

We now place ourselves in the scaling regime  where $\sigma = \chi/N$ (motivated by  numerical results). In this case the width of the distribution of $\beta-1$ scales as $N^{-1}$. Assuming that in this regime $\gamma \to 0$, one finds that to leading order $\llangle \beta^k \rrangle = m + \mathcal{O}\left( N^{-1} \right)$. Now, given the expression for $\gamma^2$ in \ref{appendix:full_equ}, one finally obtains
\begin{equation}
    \gamma = \mathcal{O} \left( {N}^{-1/2} \right), \quad \text{when} \quad \sigma = \mathcal{O} \left( {N}^{-1} \right),
\end{equation}
which justifies our assumption that $\gamma \to 0$ for large $N$. It furthermore shows that the width of the smoothed step function $F_N(\beta)$ (Eq.~\eqref{equ:F_N}) scales as $N^{-1}$.

This result then allows us to explicitly make Sommerfeld-like expansions of averages, as described in \ref{appendix:sommerfeld}, that now have no contribution at order $N^{-1}$. Eliminating the higher order terms appropriately finally yields the equation for the mean threshold
\begin{equation}
    \beta^+ = \frac{\avgc{\beta^2}}{\avgc{\beta}} + \frac{1}{N} \frac{\overline{z}}{\avgc{\beta}} + \mathcal{O} \left( \frac{1}{N^2} \right),
    \label{equ:self_cons_final}
\end{equation}
valid in the regime of interest $\sigma = \chi/N$, with $\overline{z} = {\langle z^{-1} \rangle}^{-1}$ and 
\begin{equation}
    \avgc{g(\beta)}:=\int_{-\infty}^{\beta^+} {\rm d}\beta \, g(\beta) \rho(\beta).
\end{equation} 
Eq. \eqref{equ:self_cons_final} is self-consistent in the sense that $\beta^+$ appears in both sides of the equation.
 
Recall that this equation for $\beta^+$ assumes negligible correlations between the occupation variables $\theta_i$. In the spirit of a {\it bona fide} cavity calculation, one should look at the effect of the introduction of an additional asset or species on the already existing $\theta_i$. Knowing the importance of the Onsager reaction term in Sherrington-Kirkpatrick spin-glasses, that turns the naive mean field equation into the celebrated TAP equation \cite{thouless1977solution}, it is important to ensure that the average threshold is not affected by a similar term. This is done in \ref{appendix:cavity} where  we check that the introduction of a new spin does not alter the above equation at order $N^{-1}$. As such, Eq.~\eqref{equ:self_cons_final} is our central analytic result for the problem at hand, which we shall solve for different distributions $\rho(\beta)$ in section \ref{sec:results}. 

\subsection{Complexity and number of solutions}

Now, we define $\mathcal{N}(K,N)$ to be the average number of solutions satisfying the constraint with $K$ among the $N$ possible spins included. We may write an iterative equation to describe the evolution of this quantity as $N \to N +1$. First, the addition of this new element -- that we will take to be at index 0 and associated to $\beta_0$ -- is only possible if $\beta_0$ is small enough. If we recall the probabilistic interpretation of the maximum sparsity $m = \mathrm{Prob}(\beta_0 \leq \beta^+)$, the probability of $\theta_0 = 1$ being compatible with the constraint is simply given by $m(\sigma,K)$. In order to form such a solution, with $K$ among the now $N+1$ spins included, the new element must be added to a solution previously comprising $K-1$ spins. However, a fraction of the solutions with $K-1$ nonzero spins are rendered invalid due to the fact that $\beta^+$ is a decreasing function of $K$. Those positions are such that $\beta^+(K) < \beta_i < \beta^+(K-1)$, and occur with probability
\begin{equation}
    p(\sigma,K) = \int_{\beta^+(K)}^{\beta^+(K-1)} \dd \beta \, \rho(\beta) = m(\sigma,K-1) - m(\sigma,K),
\end{equation}
and given $\beta$s are drawn independently, we finally find the expression
\begin{equation}
    \mathcal{N}(K,N+1) = \mathcal{N}(K,N) + m(\sigma,K)[1 - p(\sigma,K)]^{K-1} \mathcal{N}(K-1,N)
    \label{equ:iterative_full}
\end{equation}
to describe the evolution of the number of solutions with $K$ non-zero spins as $N$ increases. To properly initialise and close the recursion, we require
\begin{equation}
    \mathcal{N}(0,0) = 1  \quad \mathrm{and} \quad \mathcal{N}(N+1,N) = 0.
\end{equation}
The quantity that interests us, the average total number of solutions satisfying the constraint, is then simply given by
\begin{equation}
    \langle \mathcal{N}_s \rangle = \sum_{K=1}^N \mathcal{N}(K,N).
    \label{equ:sum_Ns}
\end{equation}

Defining $n(x,t)$ to be the continuous analogue of $\mathcal{N}(K,N)$ with $K \to x$ and $N \to t$, the iterative equation may be rewritten as a partial differential equation, valid in the large $N$ limit. To leading order, i.e. neglecting a diffusion term of order $N^{-1}$, one has
\begin{equation}\label{eq:advection}
    \partial_t n(x,t) + \mathrm{e}^{x\varphi'(\sigma x)} \, \varphi(\sigma x) \,  \partial_x n(x,t) = \mathrm{e}^{x\varphi'(\sigma x)} \, \varphi(\sigma x) \, n(x,t)
\end{equation}
where we have used the scaling result $m(N,\sigma) = \varphi(\chi)$ with $\chi=\sigma N$ as observed in numerical experiments, and further justified by the analytical calculations in the next section. 

This inhomogeneous advection equation may then be treated with the method of characteristics \cite{courant1962partial}. Taking the characteristic curve $s$ in ($x,t$) space, and writing $z(s) = n(x(s),t(s))$ the solution along the curve, the problem reduces to the system of ordinary differential equations
\begin{align}
    \frac{\dd t}{\dd s} &= 1 \label{equ:char1}\\
    \frac{\dd x}{\dd s} &= \mathrm{e}^{x(s)\varphi'(\sigma x(s))} \, \varphi(\sigma x(s)) \label{equ:char2}\\
    \frac{\dd z}{\dd s} &= \mathrm{e}^{x(s)\varphi'(\sigma x(s))} \, \varphi(\sigma x(s)) \, z(x(s),t(s)), \label{equ:char3}
\end{align}
with boundary conditions
\begin{equation}
    t(0) = 0, \quad x(0) = 0, \quad z(0) = 1.
\end{equation}
The solution satisfying these boundary conditions then directly corresponds, for $t=N$, to the dominating term in the sum given in Eq.~\eqref{equ:sum_Ns}.

To summarise, the self-consistent equation \eqref{equ:self_cons_final}  allows one to determine the average threshold $\beta^+$ for the inclusion of an asset or specie in the non-negative solution for a given sparsity. This quantity will in turn yield the expression of the maximum sparsity $m(N,\sigma) = \varphi(\chi)$ in the regime $\sigma = \chi/N$. Solving the set of characteristic equations tracing back to the known boundary conditions shall finally give an expression of the average number of solutions, and therefore the annealed complexity.

\section{\label{sec:results}Distribution-specific results}

\subsection{\label{subsec:Gaussian}Gaussian $\beta$}
As argued with the data presented in Fig.~\ref{fig:real_distribs}(a), taking $\beta$ to be normally distributed with mean 1 and variance $\sigma^2$ appears to be a good approximation for the portfolio problem. Going back to Eq.~\eqref{equ:self_cons_final}, all the terms of interest can be written exactly using the Gaussian cumulative distribution function $\Phi(x) = \frac{1}{2}(1+\mathrm{erf}(x/\sqrt{2}))$. Taking $\sigma = \chi/N$ and introducing the ansatz
\begin{equation}
    \beta^+ = 1 + \frac{\chi f(\chi)}{N}
\end{equation}
allows one to rewrite the self-consistent equation as
\begin{equation}
    \chi f(\chi) = \frac{\overline{z}}{m} - \frac{1}{m} \frac{\chi}{\sqrt{2\pi}} \mathrm{e}^{-\frac{1}{2} f(\chi)^2},
    \label{equ:self_cons_normal_scaled}
\end{equation}
with $m = \varphi(\chi) = \Phi(f(\chi))$. As anticipated in the previous section, $m(N,\sigma)$ indeed only depends on $\chi$ in the scaling regime.  

Setting $\overline{z} = 1$ (without loss of generality, since it simply corresponds to the rescaling $\chi \to \chi/\overline{z}$) this equation may be solved numerically for $f$ at given $\chi$, the result of which is plugged back into the expression for $m$ and is shown by the continuous line in Fig.~\ref{Fig:Gaussian}(a). This analytical result is in excellent agreement with numerical experiments, which gives us confidence that our self-consistent equation is exact in the regime of interest. As expected, while qualitatively reasonable, the model does not perfectly describe the sparsity corresponding to more broadly distributed empirical $\beta$'s.


This numerically obtained $m = \varphi(\chi)$ can also be injected in the iterative expression for $\mathcal{N}(K,N)$ given in Eq.~\eqref{equ:iterative_full}. Summing all contributions, the mean number of solutions $\langle \mathcal{N}_s \rangle$ and associated complexity $\Sigma$ are computed and shown by the continuous lines in Fig.~\ref{Fig:Gaussian}(b-c). The match between this semi-analytical solution and the numerical results is also excellent, this time for both the arbitrary and empirically determined values of~$\sigma$.

Based on the numerical solution of Eq.~\eqref{equ:self_cons_normal_scaled}, we find that $f(\chi \gg 1)$ quickly reaches large negative values. The error functions through which $\varphi$ is expressed can therefore be approximated asymptotically through the method of steepest descent. Keeping the first two terms in the series expansion of $m = \varphi$ in the self-consistent equation, and taking iterated logarithms, one finally finds, at the leading order in the scaling regime:
\begin{equation}\label{eq:varphi_asympt}
    \varphi(\chi) \approx \frac{\sqrt{2\log \chi}}{\chi}, \qquad (\chi \gg 1)
\end{equation}
This asymptotic result is compared to the numerical experiments in the inset of Fig.~\ref{Fig:Gaussian}(a), displaying a very good fit for values as small as $\chi \sim 10$.

As detailed in \ref{appendix:characteristics}, this result may be used in the characteristic Eq.~\eqref{equ:char2}. At the leading order, we find the expression of $x$ along the characteristic
\begin{equation}
    x(s) = \sqrt{\frac{2s}{\sigma}} (\log \sigma s)^\frac{1}{4}\left[1 + \mathcal{O}\left( \frac{\log \log \sigma s}{\log \sigma s} \right) \right].
\end{equation}
Eq.~\eqref{equ:char3} may then be integrated to find $z(s)$ the number of solutions along the characteristic,
\begin{equation}
    \log z(s) = \sqrt{\frac{2 s}{\sigma}} (\log \sigma s)^{\frac{1}{4}} \left[1 + \mathcal{O}\left( \frac{1}{\log \sigma s} \right) \right]
\end{equation}
From Eq.~\eqref{equ:char1} and the associated boundary condition, we may now finally set $s = t =N$. Going back to original variables of the problem, we therefore have the asymptotic evolution for the number of non-negative solutions
\begin{equation}
    \langle \mathcal{N}_s \rangle \sim \exp \left( \sqrt{\frac{2 N}{\sigma}} (\log \sigma N)^\frac{1}{4} \right)
\end{equation}
and the associated annealed complexity
\begin{equation}
    \Sigma \approx \sqrt{2} \, \frac{ (\log \chi )^{\frac{1}{4}}}{\sqrt{\chi}}.
\end{equation}
This fully analytical asymptote is compared to the previously obtained numerically exact solution of the recursion relation (dubbed ``semi-analytical'' below) in the inset of Fig.~\ref{Fig:Gaussian}(c). The result appears satisfactory, although the conclusions are limited by the numerical difficulty of obtaining the semi-analytical result for large values of $N$. A careful observation suggests a small shift between the two curves, which might be explained by the second derivative (diffusion) term in the partial differential Eq.~\eqref{eq:advection}, which we neglected. 

In any case, this result corresponds to a growth slightly faster than $\mathrm{e}^{\sqrt{N}}$ but significantly slower than $\mathrm{e}^{N}$: asymptotically, the complexity of the rank-one portfolio problem, or of the rank-one ecological problem, is zero, contrarily to the spin-glass case. But the average number of different possible solutions is still very large when $N$ is large.

It immediately appears however, that this solution is somewhat contradictory with the previously found behaviour of the maximum sparsity. Indeed, taking a closer look at the solution for $x(s)$ along the characteristic curve, we find that the associated sparsity $u^\star = x(t)/t$ is given by
\begin{equation}
    u^\star \sim \frac{(\log \chi)^\frac{1}{4}}{\sqrt{\chi}}.
\end{equation}
Comparing with Eq. \eqref{eq:varphi_asympt}, we find that $u^\star \gg \varphi(\chi)$ for $\chi \gg 1$.

In other words, it appears that the configurations dominating the count of the mean number of solutions are those with a number of non-zero spins $K^\star$ greatly exceeding the theoretical prediction $m N$.
While surprising at first, this result means that the average (over $\beta$s) number of solutions is dominated by extremely rare configurations $\{\beta_i\}$ that allow $K^\star \gg m N$ positions to survive in the portfolio. Even if rare, such configurations allow an exponentially large number of portfolios to exist, i.e. $C_N^{K^\star}$. Hence the distribution of $\mathcal{N}_s$ is heavily skewed towards large values, corresponding to events that are extremely unlikely to be witnessed in reality. For typical configurations of the $\beta_i$'s, on the other hand, one expects that eligible portfolios are much smaller, and contain at most $m N$ assets. Correspondingly, the typical number of solutions is expected to be of order of $e^{Nm} \ll \langle \mathcal{N}_s \rangle$. 
In order to compute precisely the typical number of solutions, one should compute $\langle \log \mathcal{N}_s \rangle$ and the associated quenched complexity. This would require going back to the direct formulation given in Eq.~\eqref{equ:direct}, expressing the Heaviside step function with its Fourier representation and making use of the replica trick as detailed in numerous works relating to spin-glasses (see \cite{Bray1980,Bray1981,mezard1987spin}). We leave such a calculation for later investigations. The computation of the probability of observing these rare configurations that dominate the average solution count, that is likely directly dependent on the distribution of the $\beta_i$'s, is also left for future work.

\begin{figure}
    \centering
    \includegraphics[width=0.5\linewidth]{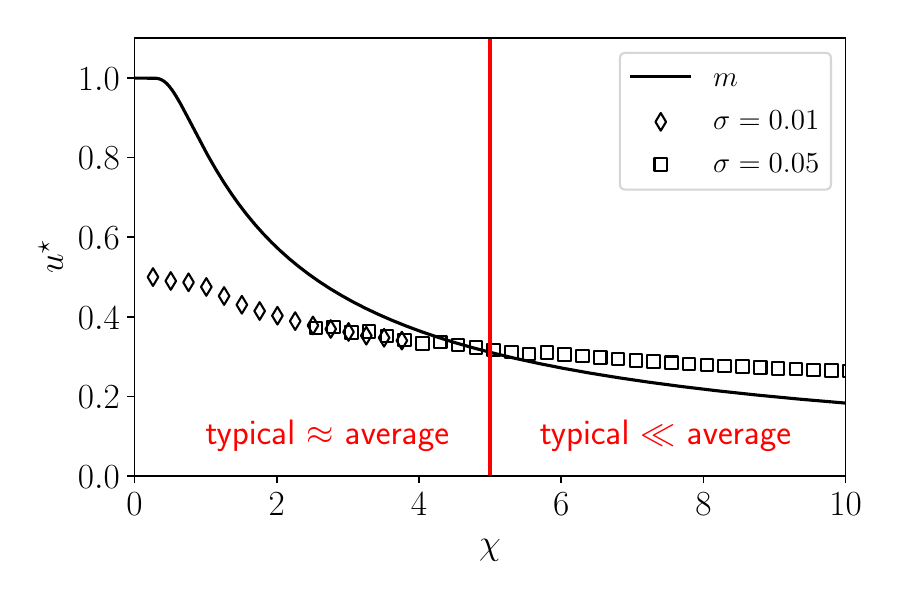}
    \captionof{figure}{Evolution of the effective sparsity of the maximum of $\mathcal{N}(K,N)$ calculated iteratively for Gaussian $\beta$ of variance $\sigma^2$, up to $N=400$ (points), compared to the maximum sparsity (line). The red vertical line separates a small $\chi$ regime where one expects that $\langle \mathcal{N}_s \rangle \approx \mathcal{N}_s^{\text{typ.}}$, from a large $\chi$ regime where $\langle \mathcal{N}_s \rangle \gg \mathcal{N}_s^{\text{typ.}}$ when $N \to \infty$.}
    \label{Fig:mean_v_typical}
\end{figure}

This difference between the typical (and average) maximum sparsity $m$ and the most likely effective sparsity $u^\star$ resulting from the iterative procedure is apparent in Fig.~\ref{Fig:mean_v_typical} that considers normally distributed $\beta$, and is comparable in the uniform case. Here, we stress that the former corresponds to the self-averaging fraction of positions that may be included in a long-only Markowitz optimal portfolio, while the later corresponds to the fraction of occupied positions associated to the portfolio that dominated the calculation of $\langle \mathcal{N}_s \rangle$ at large $N$ (i.e. the solution at the saddle in the spin-glass language). For small $\chi$, we have $u^\star = \frac{1}{2} < m$, as naively expected. However, beyond $\chi \approx 5$ we find that indeed the mean behaviour $u^\star$ exceeds the typical sparsity $m$. In this region, we would therefore expect the calculation of $\langle \mathcal{N}_s \rangle$ and the associated annealed complexity $\Sigma$ to deviate from that observed in a moderate number of numerical experiments. The divergence between $u^\star$ and $m$ appears to be relatively slow however, explaining why it is not clearly noticeable in the numerical results in Fig.~\ref{Fig:Gaussian} and \ref{Fig:uniform}, where only a small fraction of elements are excluded as $m$ is still close to unity.

\subsection{\label{subsec:Uniform}Uniform $\beta$}

We now extend the results to other distributions of $\beta$. The uniform distribution is an interesting case, as it can model the case of stocks restricted to certain industries (see~\ref{fig:real_distribs}(b)).

In the uniform case, the distribution is still centred about $\beta = 1$ but now has width $2\sqrt{2}\sigma$. Once again, we take the scaling $\chi = \sigma N$ from numerical experiments, and the ansatz $\beta^+ = 1 + \chi f(\chi)/N$. The moments up to the threshold $\beta^+$ assuming $\beta^+ < 1 + \sqrt{2}\sigma$ may then be easily written explicitly given the simple expression of the uniform distribution. Taking the self-consistent equation at order $N^{-1}$, where we can once again take $\overline{z} = 1$ without loss of generality, finally gives the expression for the function $f(\chi) = -\sqrt{2} \pm 2^{5/4}/\sqrt{\chi}$, from which the maximum sparsity directly follows by picking the solution giving the positive result. This solution requires the threshold to be before the right edge of the distribution, and hence must be completed with the result beyond which saturates the maximum value $m=1$. Combining both gives a closed form solution for the entirety of the domain without having to rely on asymptotics
\begin{equation}
    \varphi(\chi) = 
    \left\{ \begin{array}{c@{\quad}cr} 
        1 & \text{ for } \chi \leq \frac{1}{\sqrt{2}},\\
        \frac{1}{2^{1/4}\sqrt{\chi}} & \text{ for } \chi > \frac{1}{\sqrt{2}}.
    \end{array}\right.
\end{equation}
This solution corresponds to the continuous line in Fig.~\ref{Fig:uniform}(a). Once again, the match with numerical simulations is very good, whereas -- as discussed in Section~\ref{sec:numerical} -- there is a small offset relative to the empirically sampled points that lie slightly outside of the analytically tractable region. Note that the typical sparsity of the portfolios decreases with $N$ much more slowly in the uniform case than in the Gaussian case. 

The fully analytical solution for $m = \varphi(\chi)$ is substituted in the iterative formula for the mean number of solutions, resulting in the continuous lines in Fig.~\ref{Fig:uniform}(b-c). Clearly, this theoretical result displays a very good match with the numerical points across all values of $\sigma$ tested.

As before, we now employ this expression in the set of ordinary differential equations to solve the partial differential equations describing the evolution of the number of solutions. Thanks to the simple expression for $\varphi$ that is now valid for all values of $\chi$, the integration may be carried out with no difficulty (\ref{appendix:characteristics}), giving
\begin{equation}
    \log z(s) = \left( \frac{3}{2^{5/4}} \frac{s}{\sqrt{ \sigma}}\right)^\frac{2}{3} \left[1 + \mathcal{O}\left( \frac{1}{\sqrt[3]{\sigma s} } \right) \right]
\end{equation}
and thus simply replacing $s = t =N$,
\begin{equation}
    \langle \mathcal{N}_s \rangle \sim \exp \left\{ \left( \frac{3}{2^{5/4}} \frac{N}{\sqrt{ \sigma}} \right)^\frac{2}{3} \right\}.
\end{equation}
As we might have expected from the slower decrease in maximum sparsity relative to the Gaussian result, the average number of solutions grows faster in the uniform case. The annealed complexity is now asymptotically given by
\begin{equation}
    \Sigma \sim \left( \frac{3}{2^{5/4}}\frac{1}{\sqrt{\chi}} \right)^{\frac{2}{3}},
\end{equation}
that is plotted with the dashed line in the inset of Fig.~\ref{Fig:uniform}(c). Here, the fully analytical expression appears more or less in line with the semi-analytical iterative solution.

As for the Gaussian case, we notice that the sparsity of the configurations dominating the enumeration is given by $u^\star = x(t)/t \sim \chi^{-\frac{1}{3}} \gg \varphi(\chi)$. Just as before, we have therefore calculated a mean number of solutions that appears to greatly exceed the typical result observed. The typical behaviour would then also require to compute $\langle \log \Ns \rangle$, which in this uniform case would not be as similar to typical spin-glass calculations that rarely, if ever, involve uniform distributions with a finite support. We note however that the typical number of solutions in this case should grow as $\exp(\sqrt{N})$, i.e. much faster than in the Gaussian case where it only grows as $\exp(\sqrt{\log N})$.

\subsection{\label{subsec:GenNormal}Bridging the gap: generalized normal distribution}
To understand why two different decays in maximum sparsity hold for the normal and uniform distributions, we use of the generalized normal distribution
\begin{equation}
    \rho_b(\beta) = \frac{b}{2\sqrt{2} \sigma \Gamma(1/b)} \mathrm{e}^{-\big( \frac{|\beta - 1|}{\sigma \sqrt{2}} \big)^b}
\end{equation}
where $b$ is a shape parameter allowing  to recover the usual normal distribution of unit mean and standard deviation $\sigma^2$ for $b=2$, and the uniform distribution centered at 1 and of width $2\sqrt{2}\sigma$ by taking the limit $b\to \infty$. Moreover, probing $b\leq 1$ can provide insights on the problem with heavier tailed distributions of $\beta$'s, $b=1$ corresponding to the Laplace case, which may be of interest when considering e.g.~the weekly returns presented in Fig.~\ref{fig:real_distribs}(c).

The first step in our search for an analytical solution in this general formulation is to express the moments up to the threshold $\langle \beta \rangle_c$ and $\langle \beta^2 \rangle_c$ as well as $m$ itself in workable forms. As detailed in \ref{appendix:gennormal}, this can be done by reintroducing the expressions for $\sigma$ and $\beta^+$. The generalised self-consistent equation now reads
\begin{equation}
    \chi f(\chi) = \frac{\overline{z}}{m} - \frac{\chi}{m}  \int_{-f(\chi)}^\infty \mathrm{d}u \, u \,\rho(u),
\end{equation}
where $u=(\beta-1)/\sqrt{2}\sigma$ and we have postulated $f(\chi) < 0$ which is intuitive from the expression of $\beta^+$ (we expect the threshold to be smaller than the mean value of $\beta$, regardless of the distribution). Taking $b = 2$, the integral can be evaluated exactly and we recover Eq.~\eqref{equ:self_cons_normal_scaled} as expected. As for the two previous cases, setting $\overline{z}=1$ simply corresponds to rescaling $\chi \to \chi/\overline{z}$.

\begin{figure}
    \centering
    \includegraphics[width=0.58\linewidth]{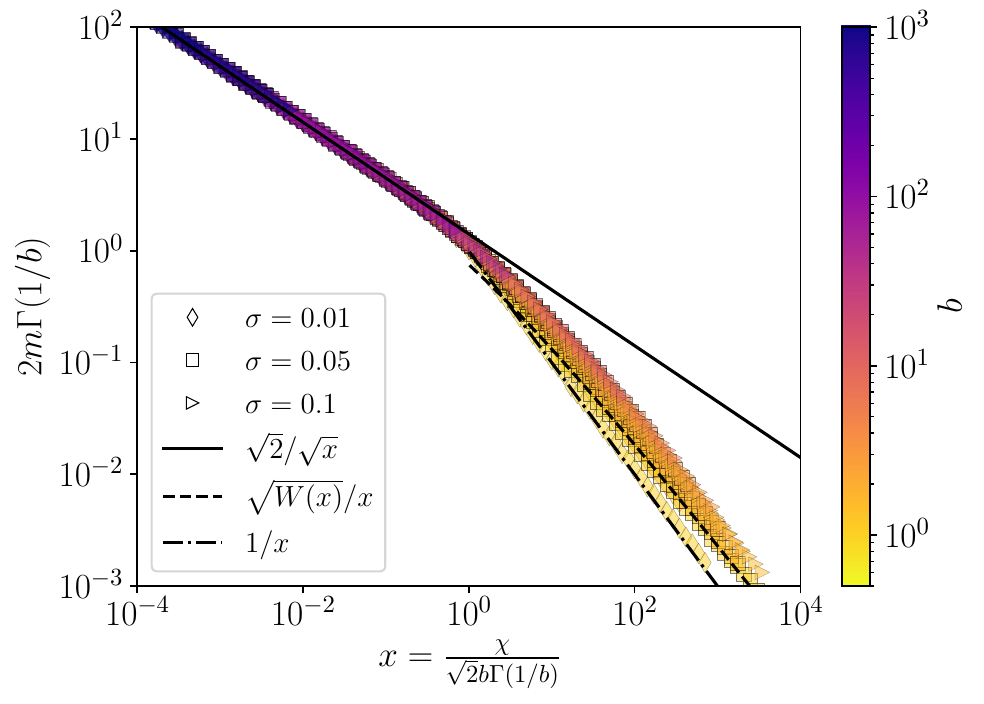}
    \captionof{figure}{Scaled sparsity as a function of $b$, $\sigma$ and $N$ for generalised normal distributions. The continuous, dashed and dot-dashed lines correspond to the uniform, Gaussian and Laplace asymptotes respectively.}
    \label{Fig:gennormal}
\end{figure}

For $b$ sufficiently small, we may approximate the integrals asymptotically as we expect $f(\chi)$ to have a large magnitude for these widely distributed $\beta$'s. The resulting self-consistent equation (\ref{appendix:gennormal}) now reads
\begin{equation}
    \mathrm{e}^{\big( \frac{|f(\chi)|}{\sqrt{2}} \big)^b} = \frac{\chi}{\sqrt{2}b \Gamma(1/b)} \left(\frac{|f(\chi)|}{\sqrt{2}} \right)^{2-2b}
\end{equation}
while the maximum sparsity at the leading order is
\begin{equation}
    m = \varphi(\chi) = \frac{\mathrm{e}^{-\big( \frac{|f(\chi)|}{\sqrt{2}} \big)^b}}{2\Gamma(1/b)} \left( \frac{|f(\chi)|}{\sqrt{2}} \right)^{1-b}.
\end{equation}
Introducing the variables 
\begin{equation}
    y = \left( \frac{|f(\chi)|}{\sqrt{2}} \right)^{b} \quad \text{and} \quad x = \frac{\chi}{\sqrt{2}b\Gamma(1/b)},
\end{equation}
the self-consistent equation takes the much simpler form
\begin{equation}
    y^{2-\frac{2}{b}}e^y = x,
    \label{equ:self_cons_xy}
\end{equation}
giving in turn $2 \Gamma(1/b) \varphi = y^{1-\frac{1}{b}}/x$. For a given value of $b$, this simplified self-consistent equation can be either solved semi-analytically or asymptotically in the limit of $\chi$ and therefore $x$ large. For instance taking $b=2$, Eq.~\eqref{equ:self_cons_xy} gives $y = W(x)$ the Lambert $W$ function and thus
\begin{equation}
    m = \frac{\sqrt{W(x)}}{2 \sqrt{\pi} x} \sim \frac{\sqrt{2 \log \chi}}{\chi},
\end{equation}
thereby recovering the previously obtained result. Interestingly, the case $b=1$ corresponding to Laplace distributed $\beta$'s yields the exact relation
\begin{equation}
    m = \frac{1}{2x} = \frac{1}{\sqrt{2} \chi},
\end{equation}
suggesting $\langle \mathcal{N}_s \rangle \sim \mathrm{e}^{\sqrt{N}}$, slightly slower than for Gaussian $\beta$'s. Both asymptotic solutions are shown in Fig.~\ref{Fig:gennormal}, displaying a good match as $\chi$ increases.

For $b \gg 1$, the problem is not as straightforward. Indeed, as the shape of the distribution approaches the uniform case, the effective support narrows to reach sharp cutoffs at $\beta = 1\pm \sqrt{2}$ for $b\to \infty$. In this limit, we therefore require $-f(\chi) < \sqrt{2}$, which is obviously unsuitable for the previously taken asymptotic approximations of the integrals. Instead, the limit $b \to \infty$ must be taken before evaluating the integral. Doing so, one can directly recover the uniform expressions for $f(\chi)$ and $m$ from the previous section, and therefore $m = (\sqrt{2} \chi)^{-\frac{1}{2}}$.

Rescaling the $b\to \infty$ result suggests $2 \varphi\Gamma(1/b) \approx \sqrt{2}/\sqrt{x}$ for large $b$.  This large $b$ solution, and the crossover between the two regimes, with   $\varphi$ decays as $\chi^{-\frac{1}{2}}$ and $\chi^{-1}$ respectively, can be seen in Fig.~\ref{Fig:gennormal}. Interestingly, for all finite values of $b$ the second regime will be reached eventually as $\chi$ is increased, and only the uniform distribution will remain in the first regime, the slower decay of which translates in a larger number of solutions. As such, the uniform distribution will be the case within the generalised normal family allowing for the largest number of solutions, with finite $b > 2$ cases only affecting the exponent of the logarithmic term in the complexity.

While the above theoretical setup should hold for finite $b < 1$, it is difficult to avoid a large number of negative $\beta$'s when considering thicker tails, at which point we would cease to observe a monotonous decrease of $m$ in $\chi$. Besides, it seems unlikely that a heavy tailed distribution of infinite support would correctly depict the distribution of asset correlations. For interacting species, negative interactions are not as unreasonable, as mutually beneficial relations between species can exist, however their study would require a different analytical framework. Yet, if negative values remain rare, it is clear from the self-consistent equation that as tails get thicker, the number of solutions $\mathcal{N}_s $ further decreases before one enters a new regime when negative $\beta$'s start proliferating.

\section{\label{sec:chaos}Disorder chaos}
Having found that the number of solutions satisfying the non-negativity constraint is near exponential for relevant distributions of $\beta$ (in the regime $\langle \beta \rangle > 0$ and $\sigma = \chi/N$), we ask ourselves if we can observe \textit{disorder chaos} in this system. Disorder chaos in this context is essentially the question of the stability of the optimal solution if the disorder $\beta$ is slightly altered, particularly in the case of large $N$. Indeed, if there is an exponential number of valid solutions, some with similar values of the objective function, it is not hard to imagine that a slight modification in the disorder could yield a complete reshuffling in the spin configuration. This idea is further supported by the numerical exploration of the solution space that was conducted in Section \ref{subsec:solution_space}, where we found that a large number of configurations with little overlap with the optimal solution indeed have very close properties.

This phenomenon has been observed in spin-glasses \cite{azcoiti1995static,Krzakala2005,monthus2014chaos}, and may be formulated in a formally very similar way. It should be noted that this form of instability under changes in the quenched disorder, sometimes also referred to as static chaos, is not to be confused with temperature chaos \cite{kondor1989chaos} (as the use of $\beta$ might induce some confusion in the spin-glass context). 
Introducing the perturbation $\varepsilon$, we alter the disorder as
\begin{equation}
    \tilde{\beta}_i - 1 = (\beta_i - 1)\left( \frac{1 + \varepsilon \upsilon_i}{\sqrt{1 + \varepsilon^2}} \right),
    \label{equ:beta_perturb}
\end{equation}
where $\upsilon_i$ is a Gaussian random variable with zero mean and unit variance. This definition allows one to keep the variance of the modified $\beta$s unchanged. 

To compare the optimal/most diverse non-negative solution to the original problem to the perturbed one, it is necessary to introduce some measure of the overlap between solutions. Recalling our definition of the overlap between two configurations, Eq. \eqref{eq:overlap},
we subsequently define the portfolio correlation as
\begin{equation}
    O_N(\mathbf{w},\tilde{\mathbf{w}}) = \frac{\langle q_{\{\theta\},\{\tilde{\theta}\}} (N) \rangle}{\sqrt{\langle q_{\{\theta\},\{\theta\}} (N)\rangle \langle q_{\{\tilde{\theta}\},\{\tilde{\theta}\}} (N) \rangle }},
\end{equation}
where  $\{ \theta \}$ and $\{ \tilde{\theta} \}$ correspond to the original and altered configurations respectively. With this definition we ensure $O_N(\mathbf{w},\tilde{\mathbf{w}}) = 0$ for independent portfolios, and $O_N(\mathbf{w},\tilde{\mathbf{w}}) = 1$ for $\mathbf{w}=\tilde{\mathbf{w}}$.

\begin{figure}
    \centering
    \includegraphics[width=0.85\linewidth]{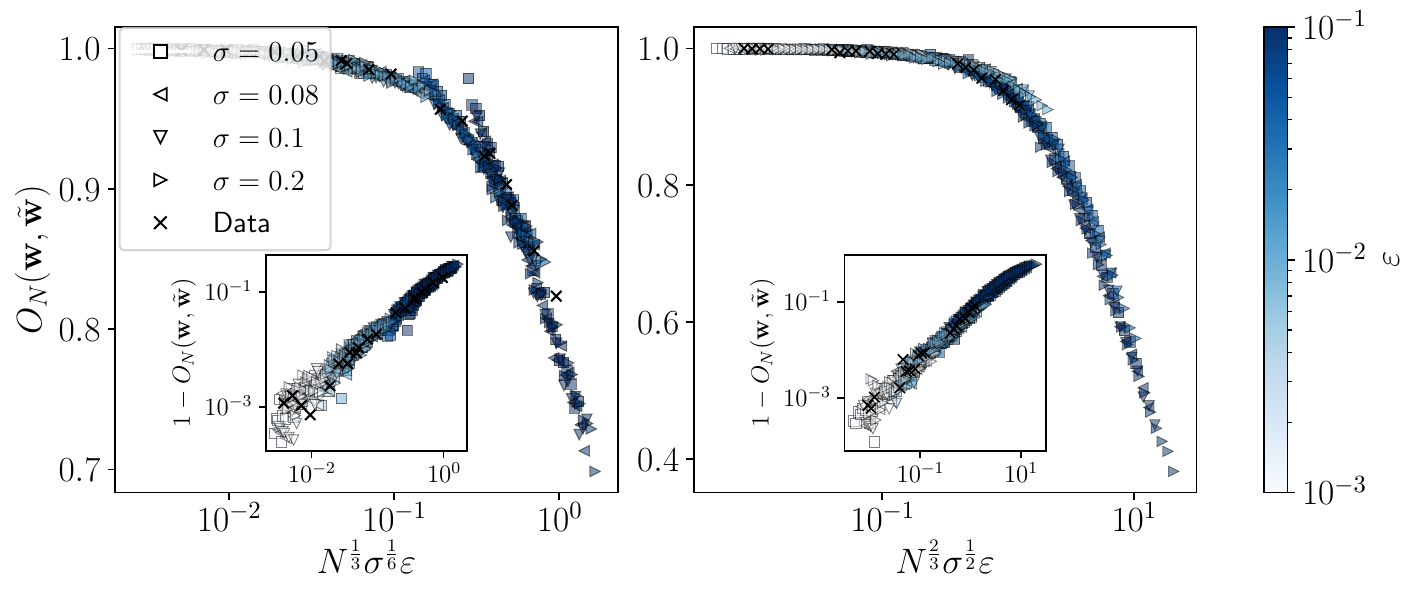}
    \caption{Overlaps obtained numerically for different fixed values of $\sigma$, $\varepsilon$ while $N$ is varied up to $10^4$, averaged over 64 realisations. Left: $\beta \sim \mathcal{N}(1,\sigma^2)$, data collapsed with $N^\frac{1}{3} \sigma^\frac{1}{6}$. Right: $\beta \sim \mathcal{U}(1\pm \sqrt{2}\sigma)$, data collapsed with $N^\frac{2}{3} \sigma^\frac{1}{2}$. Insets represents the overlap subtracted from 1, its maximum value for identical disorder, plotted in log-log.}
    \label{Fig:chaos}
\end{figure}

The resulting overlap for Gaussian and uniform $\beta$'s compatible with market data is shown in Fig.~\ref{Fig:chaos}. Qualitatively, both collapsed plots appear similar, with a decrease in the overlap as $N$ gets large and other parameters are kept fixed. Taking a closer look, it is clear that the disorder chaos is stronger for uniformly distributed $\beta$'s, which is consistent with the fact that the number of solutions $\mathcal{N}_s$ is larger in this case, as found in the previous section. While the reduction of the overlap in $N$ is easily understandable, the detailed scaling behaviour in $\sigma$ found numerically is more challenging to describe analytically.

In any case, the logarithmically scaled insets shows a clean power law behaviour in $N$. Asymptotically, our results therefore suggest that $O_N(\mathbf{w},\tilde{\mathbf{w}}) \to 0$ for any $\varepsilon > 0$ as $N \to \infty$, characteristic of disorder chaos. On both plots, the data points sampled from empirical $\beta$'s are well aligned with numerical simulations, suggesting the chaos observed is robust somewhat beyond the regime studied analytically. 

Regardless of the precise behaviour of the overlap with the problem's parameters, the disorder chaos observed here is first and foremost a qualitative insight. What this observation tells us has a practical consequence on the way one might approach the systematic construction of a long-only portfolio. Supposing one picks among 3000 stocks for example, which is a reasonable number for a large asset manager, a change of the order of 10\% of the $\beta$'s could result in a significant reshuffling in the positions that should be held, particularly if the assets considered are within the same or similar industries for instance. Such a change could e.g.~come from empirical estimation errors of the correlations, or simply because the $\beta$'s naturally evolve in time depending on the many factors not incorporated in the present risk model. Besides, if one decides to modify the portfolio to match the new optimal result, it is likely that significant transaction costs could come into play, particularly given the highly concentrated nature of large long-only portfolios, so it would rather make sense to choose a portfolio that is a mix of many different quasi-degenerate solutions of the optimisation problem. Note that such a portfolio would then not be optimal in the Markowitz sense (i.e.~would not satisfy Eq.~\eqref{equ:Markowitz}), but would likely reduce the volatility as the $\beta_i$'s are allowed to vary.

From a conceptual point of view, disorder chaos means that two perfectly rational investors with a slightly different method of estimating the $\beta$s might end up with very different optimal solutions in the large $N$ limit. As emphasised in \cite{Galluccio1998} and recalled in the introduction, the presence of a very large number of quasi-degenerate solutions, at the heart of disorder chaos, leads to some irreducible uncertainty in the decision of agents, even assumed to be fully rational.  

In terms of ecological equilibria, this observation also has concrete implications. Indeed, it suggests that a moderate change in the interaction between the $N$ species considered can lead to a significantly different outcome in terms of surviving species at the equilibrium. It seems reasonable to imagine that some physical changes to the environment (e.g. through temperature changes or the introduction of chemicals) could alter the strength of interactions between species, which could then lead to a significantly different equilibrium picture of the ecosystem (on this point, see also \cite{biroli2018marginally}).
 
\section{\label{sec:conclusion}Conclusion}

Let us summarise what we have achieved in this study. Through the introduction of a spin-glass inspired formalism, we have shown that $N$ assets or species can be recombined in a exponential number of solutions satisfying the non-negativity constraint associated to the portfolio and ecological equilibrium problems, in the special case where the interaction matrix is of unit rank. More precisely, we have computed the average (or annealed) number of solutions and have shown that its logarithm grows as $N^\alpha$, where $\alpha \leq 2/3$ depends on the distribution of asset correlations and interaction strength respectively. This average number does however not correspond to the typical behaviour of the system, observed through a limited number of numerical experiments for example. Indeed, we have found that in this problem the mean number of solutions is heavily skewed by the existence of very unlikely occurrences that yield an exponential number of solutions. Finding the typical (or quenched) number of solutions, by means of a replica calculation, therefore appears to be a natural extension of the present work. We conjecture that the result will be related to the typical sparsity $m(N)$ of the solutions, namely $\langle \log \mathcal{N}_s \rangle \propto N m(N)$. Hence, the number of possible long-only configurations that can be constructed from the $N$ entities considered remains large, specially for a strictly bounded distributions of $\beta$'s for which $N m(N) \sim \sqrt{N}$.


We have also shown numerically that the solution landscape is similar to that of other complex optimisation problems like spin-glasses, i.e. many very different configurations or portfolios are quasi-degenerate, in the sense that they lead to nearly identical values of the objective function (energy for spin-glasses, risk for portfolios). Correspondingly, the phenomenon of ``disorder chaos'' in spin-glasses, i.e. the extreme sensitivity of the optimal solution on the detailed specification of the problem when $N$ is large, is also present in our long-only portfolio problem (or in its ecological counterpart).

For asset management, this result suggests that, in the presence of transaction costs, the construction of long-only portfolios should account for such an instability and in fact blend together optimal portfolios obtained by slightly varying the risk model (here the value of the $\beta$'s). Likewise, as emphasised in the original paper from Gallucio \textit{et al.}~\cite{Galluccio1998} and recalled above, such a sensitivity is interesting in the sense that it questions the meaning of a rational decision when there is a very large number of quasi-degenerate (or ``satisficing'') solutions.


For ecological equilibria, while there is unfortunately no empirical data to support our choice of interaction matrix and to choose appropriate distributions of $\beta$, we believe that most of the conclusions drawn for parameters compatible with stocks should hold for highly competitive environments with a large number of similarly interacting species, as discussed in a different context in \cite{biroli2018marginally}. Indeed, the analytical description can be generalised to any values of $\langle \beta \rangle > 0$ that could be appropriate for the ecology problem, and we have shown that our results are in fact valid for a wide range of distributions of $\beta$.

While not explicitly discussed in the bulk of the paper, heterogeneous expected returns (or growth rates) $\mu_i$ can be analysed similarly, see \ref{appendix:heterogeneous}. We find that the solution is akin to the one obtained with $\mu_i \equiv 1$, with a threshold that is no longer on $\beta$ alone but on the ratio $\beta/\mu$. 

In both the portfolio and population dynamics cases, the choice of the effective interaction matrix $C_{ij}$ is the main limiting factor in our study. Extending results to more general (random) matrix models could be an interesting avenue to explore in the future. This being said, the very general formulation of the problem, in essence studying the non-negativity of a linear equation, leads us to believe that long-only portfolios and ecological equilibria are not the only applications for the analytical description detailed in Section \ref{sec:self-cons}. Due to its links with population dynamics, the survival of firms in macroeconomic systems \cite{PhysRevE.100.032307,dessertaine2020t} could for example be another problem to study with this spin-glass inspired approach. It has finally been brought to our attention that the study of optimal strategies in matrix games, where  mixed strategies are represented by non-negative vectors satisfying a linear equation and the fraction of strategies included in the solution is a key metric \cite{berg1998matrix}, could also be treated very similarly under appropriate assumptions regarding the payoff matrix.

\section*{\label{sec:merci}Acknowledgements}
We deeply thank 
Giulio Biroli, Guy Bunin, Th\'eo Dessertaine, Samy Lakhal, Charles-Albert Lehalle, Iacopo Mastromatteo and Jos\'e Moran for fruitful discussions, and Stanislao Gualdi for providing the data. This research was conducted within the Econophysics \& Complex Systems Research Chair, under the aegis of the Fondation du Risque, the Fondation de l’Ecole polytechnique, the Ecole polytechnique and Capital Fund Management. 


\appendix

\section{\label{appendix:full_equ}Full self-consistent equation}
Starting from
\begin{equation}
    \tilde{\beta}^+ = \frac{\sum_j \beta_j^2 \theta_j/z_j + 1}{\sum_j \beta_j \theta_j/z_j}
    \label{equ:appendix_beta_init}
\end{equation}
we make use of the central limit theorem as for $N\gg 1$
\begin{equation}
    \frac{1}{N} \sum_j \beta^k_j \theta_j /z_j \simeq \langle z^{-1} \rangle \llangle \beta^k \rrangle + \frac{1}{\sqrt{N}} \xi_k
\end{equation}
where $\xi_k$ are Gaussian noises with mean $\langle \xi_k \rangle$ = 0 and variance
\begin{equation}
    \langle \xi_k^2 \rangle = \langle z^{-2} \rangle \llangle \beta^{2k} \rrangle - \langle z^{-1}\rangle^2 \llangle \beta^k \rrangle^2.
\end{equation}
After factorisation and expansion of the denominator, Eq.~\eqref{equ:appendix_beta_init} can be written as
\begin{equation}
\begin{aligned}
    \tilde{\beta}^+ =& \frac{\llangle \beta^2 \rrangle}{\llangle \beta \rrangle} + \frac{1}{N}\left[ \frac{\overline{z}}{\llangle \beta \rrangle} - \frac{\overline{z}^2}{\llangle \beta \rrangle^2}\left(\langle \xi_1 \xi_2 \rangle -\frac{\llangle \beta^2 \rrangle}{\llangle \beta \rrangle} \langle \xi_1^2 \rangle \right) \right] + \mathcal{O}\left( \frac{1}{N^2} \right) \\
    &+ \underbrace{\frac{1}{\sqrt{N}} \frac{\overline{z}}{\llangle \beta \rrangle} \left( \xi_2 - \frac{\llangle \beta^2 \rrangle}{\llangle \beta \rrangle} \xi_1 \right) + \mathcal{O}\left( \frac{1}{N^{3/2}}. \right)}_{\mathrm{fluctuations}},
\end{aligned}
\end{equation}
with $\overline{z} = \langle z^{-1} \rangle^{-1}$, which can be rewritten as $\tilde{\beta}^+ = \beta^+ + \frac{1}{\sqrt{N}} \xi$ with the final noise term
\begin{equation}
    \xi = \frac{\overline{z}}{\llangle \beta \rrangle} \left( \xi_2 - \frac{\llangle \beta^2 \rrangle}{\llangle \beta \rrangle} \xi_1 \right)
\end{equation}
that still has zero mean and variance
\begin{equation}
    \gamma^2 = \frac{\overline{z}^2 \langle z^{-2} \rangle}{\llangle \beta \rrangle^2}\left( \llangle \beta^4 \rrangle - 2 \frac{\llangle \beta^2 \rrangle \llangle \beta^3 \rrangle}{\llangle \beta \rrangle}+ \left( \frac{\llangle \beta^2 \rrangle}{\llangle \beta \rrangle} \right)^2 \llangle \beta^2 \rrangle \right).
\end{equation}
Substituting the correct values for $\langle \xi_1 \xi_2 \rangle$ and $\langle \xi_1^2 \rangle$, the deterministic term can be rewritten as
\begin{equation}
    \beta^+ = \frac{\llangle \beta^2 \rrangle}{\llangle \beta \rrangle} + \frac{1}{N}\left[ \frac{\overline{z}}{\llangle \beta \rrangle} - \frac{\overline{z}^2 \langle z^{-2} \rangle}{\llangle \beta \rrangle^2}\left( \llangle \beta^3 \rrangle - \frac{\llangle \beta^2 \rrangle^2}{\llangle \beta \rrangle} \right) \right] + \mathcal{O}\left( \frac{1}{N^2} \right).
    \label{equ:mean_selfcons_full}
\end{equation}
Now, as detailed in the following section for the case $\alpha = 1$, at the leading order one may Taylor expand the averages about the threshold $\beta^+$ as $F_N(\beta) = H(N^\alpha (\beta-\beta^+))$, with $\alpha \geq 1/2$ from the form of $F_N$ found in the main text. As a result, $\llangle \beta^k \rrangle \simeq m + \sigma(\dots) + \dots + \sigma^k(\dots)$ and so in the regime $\sigma = \chi/N$ we have both
\begin{equation*}
    \llangle \beta^3 \rrangle - \frac{\llangle \beta^2 \rrangle^2}{\llangle \beta \rrangle} = 0 +\mathcal{O}\left( \frac{1}{N} \right) \quad \mathrm{and} \quad \llangle \beta^4 \rrangle - 2 \frac{\llangle \beta^2 \rrangle \llangle \beta^3 \rrangle}{\llangle \beta \rrangle}+ \left( \frac{\llangle \beta^2 \rrangle}{\llangle \beta \rrangle} \right)^2 \llangle \beta^2 \rrangle = 0 + \mathcal{O}\left( \frac{1}{N} \right).
\end{equation*}
As such, the second term in $N^{-1}$ may be eliminated from Eq.~\eqref{equ:mean_selfcons_full}, and we find that the variance $\gamma^2$ is at most of order $N^{-1}$.

\section{\label{appendix:sommerfeld}Sommerfeld-like expansions}
Starting from the averages
\begin{equation}
    \llangle \beta^k \rrangle = \int_{-\infty}^\infty\mathrm{d}\beta \, \beta^k \rho(\beta) F_N(\beta),
\end{equation}
we may define $\psi(\beta) = \int_{-\infty}^\beta \mathrm{d}\beta \, \beta^k \rho(\beta)$ such that integrating by parts
\begin{equation}
    \int_{-\infty}^\infty\mathrm{d}\beta \, \beta^k \rho(\beta) F_N(\beta) =  - \int_{-\infty}^\infty\mathrm{d}\beta \, \psi(\beta) F_N'(\beta).
\end{equation}
as the boundary term vanishes given $\psi(-\infty)\rho(-\infty) = 0$ and $F_N(+\infty) = 0$. As previously mentioned, $F_N'(\beta)$ is peaked in a small region around $\beta^+$, therefore we can expand $\psi$ in this region with a Taylor series, giving in turn
\begin{equation}
\begin{aligned}
    \int_{-\infty}^\infty\mathrm{d}\beta \, \varphi(\beta) F_N(\beta) =& -\psi(\beta^+) \int_{-\infty}^\infty \mathrm{d}\beta \, F_N'(\beta) \\
    &- \psi'(\beta^+) \int_{-\infty}^\infty \mathrm{d}\beta \, (\beta-\beta^+) F_N'(\beta) + \mathcal{O}((\beta-\beta^+)^2).
\end{aligned}
\end{equation}
Now, changing variables as $x = N (\beta-\beta^+)$ the equation becomes
\begin{equation}
    \int_{-\infty}^\infty\mathrm{d}\beta \, \varphi(\beta) F_N(\beta) = -\psi(\beta^+) \int_{-\infty}^\infty \mathrm{d}x \, H'(x) - \frac{1}{N} \psi'(\beta^+) \int_{-\infty}^\infty \mathrm{d}x \, x H'(x) + \mathcal{O}(N^{-2})
\end{equation}
where the first integral in $x$ easily gives -1 given our knowledge of $H(x)$, while the second can be written as 
\begin{equation}
    \kappa = -\int_{-\infty}^\infty \mathrm{d}x \, x H'(x) = \int_0^{\infty} \mathrm{d}x \, (H(x) + H(-x) -1)
\end{equation}
that will clearly be zero in the case of a symmetric function written like $H(x>0) = \frac{1}{2}+\epsilon(x)$ and $H(x<0) = \frac{1}{2}-\epsilon(x)$. Given that in our case $H(x)$ is a complementary error function and is thus symmetric, we therefore have
\begin{equation}
    \intR \dd \beta \beta^k \rho(\beta) F_N(\beta) = \int_{-\infty}^{\beta^+} \mathrm{d}\beta \, \beta^k \rho(\beta) + \mathcal{O}\left( \frac{1}{N^2} \right)
\end{equation}
or, using the compact notations introduced in the main text,
\begin{equation}
    \llangle \beta^k \rrangle = \avgc{\beta^k} + \mathcal{O}\left(  \frac{1}{N^2} \right)
\end{equation}

\section{\label{appendix:cavity}Reaction term}
We essentially adapt the Onsager cavity field approach to our problem. Considering a system with $N$ assets and their associated spins $\{ \theta \}$, the threshold for inclusion was shown to be given by 
\begin{equation}
    \tilde{\beta}^+_N = \frac{\sum_j \beta_j^2 \theta_j/z_j + 1}{\sum_j \beta_j \theta_j/z_j} = \beta^+_N + \frac{1}{\sqrt{N}} \xi
\end{equation}
After introduction of a new asset, at the index 0 for simplicity, this threshold is altered as
\begin{equation}
    \tilde{\beta}^+_{N+1} = \frac{\sum_j \beta_j^2 \theta_j/z_j + \beta_0^2 \theta_0/z_0 + 1}{\sum_j \beta_j \theta_j/z_j + \beta_0 \theta_0/z_0}
\end{equation}
that can be expressed, after applying the central limit theorem to sums and expanding the denominator as before, as
\begin{equation}
    \tilde{\beta}^+_{N+1} = \beta^+_N + \frac{1}{\sqrt{N}} \xi + \frac{1}{N} c(\beta_0)
\end{equation}
with the reaction term
\begin{equation}
    c(\beta_0) = \frac{1}{\llangle \beta \rrangle} \left(\frac{\beta_0^2 \theta_0}{z_0} - \frac{\llangle \beta^2 \rrangle}{\llangle \beta \rrangle} \frac{\beta_0 \theta_0}{z_0} \right).
    \label{equ:appendix_reaction_field}
\end{equation}
Now, just like we took $F_N(\beta)  = \mathrm{Prob}\left( \xi \geq \beta - \beta^+ \right)$, we have $F_{N+1}(\beta)  = \mathrm{Prob}\left( \xi \geq \beta - \beta^+  - \frac{1}{N} c(\beta_0) \right)$, which may be Taylor expanded and averaged over the distribution $\beta_0$ such that
\begin{equation}
    F_{N+1}(\beta) = F_N(\beta) + \frac{1}{N} \frac{\mathrm{e}^{-\frac{1}{2}\big(\frac{\beta - \beta^+_N}{\gamma} \big)^2}}{\sqrt{2\pi \gamma^2}} \intR \dd \beta_0 \, c(\beta_0) \rho(\beta_0) F_{N}(\beta_0) + \mathcal{O}\left( \frac{1}{N^2} \right).
\end{equation}
Then, simply going back to the definition of the averages $\llangle \beta^k \rrangle$, we clearly find the the integral in the second term becomes
\begin{equation}
    \llangle c(\beta_0) \rrangle = \frac{1}{\llangle \beta \rrangle z_0} \left(\llangle \beta^2 \rrangle - \frac{\llangle \beta^2 \rrangle}{\llangle \beta \rrangle} \llangle \beta \rrangle \right) = 0.
\end{equation}
As such, the reaction term has no contribution at order $1/N$, and the naive self-consistent equation for $\beta^+$ requires no further modification.

\section{\label{appendix:characteristics}Detailed resolution of the characteristic equations}
Starting with the Gaussian case, the characteristic equation for $x$, rewritten as
\begin{equation}
    s = \int_0^x \dd v \, \frac{\mathrm{e}^{-v \varphi'(\sigma v)}}{\varphi(\sigma v)}
    \label{equ:char2_int}
\end{equation}
is split between the small $\sigma v$ region, that yeilds a constant contribution, and the large $\sigma v$ region where we had the asymptote
\begin{equation}
    \varphi(\sigma v) = \frac{\sqrt{2\log \sigma v}}{\sigma v}.
\end{equation}
Using this expression, we explicitely write the derivative in the exponent
\begin{equation}
    \varphi'(\sigma v) = \frac{1 - 2 \log \sigma v}{\sigma v^2 \sqrt{2 \log \sigma v}}.
\end{equation}
Now, for $v \gg 1$, $v\varphi'(\sigma v)$ decreases like $\sqrt{\log \sigma v}/(\sigma v) \gg 1$, justifying a Taylor expansion of the exponential. As such,
\begin{equation}
    \mathrm{e}^{-v\varphi'(\sigma v)} = 1 + \frac{2\log \sigma v - 1}{\sigma v \sqrt{2\log \sigma v}} + \mathcal{O}\left( \frac{\log \sigma v}{(\sigma v)^2} \right)
\end{equation}
and thus the integrand of Eq.~\eqref{equ:char2_int} is given by
\begin{equation}
    \frac{\mathrm{e}^{-v \varphi'(\sigma v)}}{\varphi(\sigma v)} = \frac{\sigma v}{\sqrt{2 \log \sigma v}} + 1 - \frac{1}{2\log \sigma v} + \mathcal{O}\left( \frac{\sqrt{\log \sigma v}}{\sigma v} \right).
    \label{equ:integrand_final}
\end{equation}
The integration of the first term presents a slight challenge, but taking the change of variable $\sigma v = \mathrm{e}^{\frac{w^2}{2}}$,
\begin{equation}
    \int^x \dd v \, \frac{\sigma v}{\sqrt{2 \log \sigma v}} = \frac{1}{\sigma}\int^{\sqrt{2 \log \sigma x}} \dd w \, \mathrm{e}^{w^2} = \frac{1}{\sigma} \mathrm{e}^{2 \log \sigma x} F \left(\sqrt{2 \log \sigma x} \right)
\end{equation}
with $F$ the Dawson integral function. Using the asymptotic expansion of this special function \cite{spanier1987atlas}, this first term finally becomes
\begin{equation}
    \frac{\sigma x^2}{2\sqrt{2\log \sigma x}}\left[  1 + \mathcal{O}\left(  \frac{1}{\log \sigma x}\right) \right].
\end{equation}
The third term in Eq.~\eqref{equ:integrand_final} can also be expanded asymptotically, as
\begin{equation}
    \frac{1}{2} \int^x \dd v \, \frac{1}{\log \sigma v} = \frac{\mathrm{li}(\sigma x)}{2\sigma} = \frac{x}{2\log \sigma x}\left[ 1 + \mathcal{O}\left(  \frac{1}{\log \sigma x}\right) \right]
\end{equation}
where equalities are up to an additive constant, and $\mathrm{li}$ is the well known logarithmic integral function. Bringing everything together,
\begin{equation}
    s = \frac{\sigma x^2}{2\sqrt{2\log \sigma x}}\left[  1 + \mathcal{O}\left(  \frac{1}{\log \sigma x}\right) \right] + x\left[1 - \frac{1}{2\log \sigma x} + \mathcal{O}\left( \frac{1}{(\log \sigma x)^2} \right) \right] + \mathcal{O}\left( (\log \sigma x)^\frac{3}{2} \right).
\end{equation}
In the large $x$ limit, the very first term will largely dominate others and we may therefore recover the expression given in the main text,
\begin{equation}
    s \sim \frac{\sigma x^2}{2 \sqrt{2 \log \sigma x}}.
\end{equation}
Rearranging this expression, we have
\begin{equation}
    \sigma x = \sqrt{2\sigma s }(2 \log \sigma x)^\frac{1}{4}
\end{equation}
and thus we can take the iterated logarithm
\begin{equation}
    \log \sigma x = \frac{1}{2} \log \sigma s + \frac{3}{4} \log 2 + \frac{1}{4} \log \log \sigma x
\end{equation}
and thus
\begin{equation}
    \sigma x = \sqrt{2 \sigma s} (\log \sigma s)^\frac{1}{4} \left[1 + \mathcal{O}\left( \frac{\log \log \sigma s}{\log \sigma s} \right)  \right]^\frac{1}{4},
\end{equation}
giving the asymptotic relation
\begin{equation}
    x(s) \sim \sqrt{\frac{2s}{\sigma}}(\log \sigma s)^\frac{1}{4}.
\end{equation}

The final characteristic ODE can then be integrated,
\begin{equation}
    \log z = \int_0^s \dd v \, \mathrm{e}^{x(v) \varphi'(\sigma x(v))}\, \varphi(\sigma x(v)),
    \label{equ:char3_int}
\end{equation}
once again splitting the constant contribution from the small $\sigma v$ region and the known asymptotic behaviour. Replacing with the expression for $x(s)$, we have
\begin{equation}
    \log \sigma x(v) = \frac{1}{2} \log \sigma v \left[ 1 + \mathcal{O}\left( \frac{\log \log \sigma v}{\log \sigma v} \right) \right]
\end{equation}
and thus
\begin{equation}
    x(v) \varphi'(\sigma x(v)) = -\frac{(\log \sigma v)^{\frac{1}{4}}}{\sqrt{\sigma v}} \left[ 1 + \mathcal{O}\left( \frac{\log \log \sigma v - 1}{\log \sigma v} \right) \right].
\end{equation}
For $\sigma v \gg 1$, the exponential term in Eq.~\eqref{equ:char3_int} can therefore be Taylor expanded. Given 
\begin{equation}
    \varphi(\sigma x(v)) = \frac{(\log \sigma v)^{\frac{1}{4}}}{\sqrt{2 \sigma v}}\left[ 1 + \mathcal{O}\left( \frac{\log \log \sigma v}{\log \sigma v} \right) \right],
\end{equation}
the integral can finally be written as
\begin{equation}
    \log z = \int^s \dd v \, \frac{(\log \sigma v)^{\frac{1}{4}}}{\sqrt{2 \sigma v}} - \int^s \dd v \, \frac{\sqrt{\log \sigma v}}{\sqrt{2} \sigma v} + \mathcal{O}\left( \int^s \dd v \, \frac{\log \log \sigma v}{\sqrt{\sigma v} (\log \sigma v)^{\frac{3}{4}}} \right).
\end{equation}
Now, the first term will clearly dominate for large $s$. The integral may be evaluating by taking the change of variable $\sigma v = \mathrm{e}^{2w}$:
\begin{equation}
    \int^s \dd v \, \frac{(\log \sigma v)^{\frac{1}{4}}}{\sqrt{2\sigma v}} = \frac{2^{\frac{3}{4}}}{\sigma} \int^{\frac{1}{2}\log \sigma s} \dd w \, w^{\frac{1}{4}} \mathrm{e}^w = \sqrt{\frac{2s}{\sigma}} (\log \sigma s)^{\frac{1}{4}} \left[ 1 + \mathcal{O}\left( \frac{1}{\log \sigma s}  \right) \right]
\end{equation}
where the final equality may be shown by integrating by parts \cite{moll2015special}.

So far, we have used the position along the characteristic $s$, however integrating the first characteristic equation with the associated boundary condition, we have $s = t$. Recalling that the continuous variable $t$ is analogous to the size of the problem $N$ and $z$ to the average number of solutions $\langle \Ns \rangle$, we may finally express the result with the quantites of interest
\begin{equation}
    \langle \Ns \rangle \sim \exp \left\{ \sqrt{\frac{2N}{\sigma}}(\log \sigma N)^{\frac{1}{4}} \right\},
\end{equation}
as given in the main text. It should be noted that as the error terms are of logarithmic orders, we expect the convergence to this asymptote to be relatively slow in $N$.

For the uniform case, the calculations are much easier thanks to the simpler form of the maximum sparsity. For $\sigma v > 2^{-\frac{1}{2}}$ large we remind that
\begin{equation}
    \varphi(\sigma v) = \frac{1}{2^{\frac{1}{4}} \sqrt{\sigma v}},
\end{equation}
giving in turn
\begin{equation}
    v\varphi'(\sigma v) = -\frac{1}{2^{\frac{5}{4}} \sqrt{\sigma v}}.
\end{equation}
The integral given in Eq.~\eqref{equ:char2_int} therefore amounts to
\begin{equation}
    s = \int^s \dd v \, \left[ 2^{\frac{1}{4}} \sqrt{\sigma v} + \frac{1}{2} + \mathcal{O}\left( \frac{1}{\sqrt{\sigma v}} \right)\right] 
\end{equation}
after Taylor expanding the exponential term, and removing the constant contribution by using the boundary condition. One then easily finds
\begin{equation}
    s = \frac{2^{\frac{5}{4}} \sqrt{\sigma} x^{\frac{3}{2}}}{3} \left[ 1 + \mathcal{O}\left(  \frac{1}{\sqrt{\sigma x}} \right) \right],
\end{equation}
and thus for $\sigma x \gg 1$, which is expected for large $N$,
\begin{equation}
    x(s) \sim \left( \frac{3}{2} \frac{s}{\sqrt{\sqrt{2}\sigma}} \right)^{\frac{2}{3}}.
\end{equation}

As for the Gaussian case, this may be reinjected in the expressions of $\varphi$ and $\varphi'$ to calculate $\log z$. We find
\begin{equation}
    \varphi(\sigma x(v)) = \left( \frac{2}{3} \frac{1}{\sqrt{2} \sigma v} \right)^{\frac{1}{3}} \quad \text{and} \quad x(v) \varphi'(\sigma x(v)) = -\frac{1}{2} \left( \frac{2}{3} \frac{1}{\sqrt{2} \sigma v} \right)^{\frac{1}{3}},
\end{equation}
resulting in, after Taylor expanding the exponential term, 
\begin{equation}
\begin{aligned}
    \log z &= \int^s \dd v \, \left( \frac{2}{3} \frac{1}{\sqrt{2} \sigma s} \right)^{\frac{1}{3}} + \frac{1}{2} \int^s \dd v \, \left( \frac{2}{3} \frac{1}{\sqrt{2} \sigma s} \right)^{\frac{2}{3}} + \mathcal{O}\left( \int^s \dd v \, \frac{1}{\sigma v} \right) \\
    &= \left( \frac{3}{2} \frac{s}{\sqrt{\sqrt{2} \sigma }} \right)^{\frac{2}{3}} \left[ 1 + \mathcal{O}\left(  \frac{1}{\sqrt[3]{\sigma s}}\right) \right].
\end{aligned}
\end{equation}
Like before, realising $s = t$ directly gives the mean number of solutions and associated annealed complexity as a function of $N$.

\section{\label{appendix:gennormal}Generalized normal distribution}
Taking the generalised normal distribution and performing the change of variable $u = \frac{\beta-1}{\sqrt{2}}$, we have
\begin{equation}
m = \frac{b}{2\sigma \Gamma(1/b)} \int_{-\infty}^{\frac{\beta^+ - 1}{\sqrt{2}}} \dd u \, \mathrm{e}^{-\big( \frac{|u|}{\sigma} \big)^{b}},
\end{equation}
\begin{equation}
\langle \beta \rangle_c = m + \frac{b\sqrt{2}}{2\sigma \Gamma(1/b)} \int_{-\infty}^{\frac{\beta^+ - 1}{\sqrt{2}}} \dd u \, u \, \mathrm{e}^{-\big( \frac{|u|}{\sigma} \big)^{b}}
\end{equation}
and finally
\begin{equation}
\begin{aligned}
\langle \beta^2 \rangle_c &=  m + \frac{b\sqrt{2}}{\sigma \Gamma(1/b)} \int_{-\infty}^{\frac{\beta^+ - 1}{\sqrt{2}}} \dd u \, u \, \mathrm{e}^{-\big( \frac{|u|}{\sigma} \big)^{b}} \\
&+ \frac{b}{\sigma \Gamma(1/b)} \int_{-\infty}^{\frac{\beta^+ - 1}{\sqrt{2}}} \dd u \, u^2 \, \mathrm{e}^{-\big( \frac{|u|}{\sigma} \big)^{b}}.
\end{aligned}
\end{equation}
At this stage, one may first notice that if $\sigma \sim N^{-1}$, then the integrals involving $u^k$ will be of order $N^{-k}$. As such, as we are interested only in terms in $N^{-1}$ or higher, rewriting $m = \varphi$ and $\langle \beta \rangle_c = \varphi - \psi$ we therefore also have $\langle \beta^2 \rangle_c = \varphi - 2 \psi + \mathcal{O}(N^{-2})$, where $\psi$ scales as $N^{-1}$. As such, the self-consistent equation simplifies to
\begin{equation}
    \beta^+ = \frac{\varphi - 2\psi}{\varphi - \psi} + \frac{1}{N} \frac{\overline{z}}{\varphi} + \mathcal{O}\left( \frac{1}{N^2} \right).
\end{equation}
At this stage, we can reintroduce the ansatz $\beta^+ = 1+ \chi f(\chi)/N$ such that at order $N^{-1}$ the self-consistent equation becomes
\begin{equation}
    \chi f(\chi) = \frac{\overline{z}}{\varphi} - \frac{N\psi}{\varphi}
    \label{equ:self_cons_appendix}
\end{equation}
that corresponds to the equation given in the main text
\begin{equation}
    \chi f(\chi) = \frac{\overline{z}}{m} - \frac{\chi}{m} \frac{b}{2\sqrt{2}\Gamma(1/b)} \int_{-f(\chi)}^\infty \mathrm{d}u \, u \,\mathrm{e}^{-\big( \frac{|u|}{\sqrt{2}} \big)^b}.
\end{equation}

We now look at the case of finite $b$. The know asymptote
\begin{equation}
\int_{a}^{b} \dd t \, f(t) \, \mathrm{e}^{x t} \sim \mathrm{e}^{x b}\left[\sum_{k=1}^{n}(-1)^{k-1} f^{(k-1)}(b) x^{-k}\right]
\end{equation}
as $x \to \infty$ can then be used with minor tweaking. For all three integrals, we can use the fact that in the regime of interest $\beta^+ - 1 < 0$ so being careful with signs one may take the substitution $t = u^b$ to have an integrand in the form of the formula above. This then yields
\begin{equation}
    \varphi \simeq \frac{1}{2\Gamma(1/b)} \mathrm{e}^{-\big( \frac{|f(\chi)|}{\sqrt{2}} \big)^b} \left[\left(\frac{|f(\chi)|}{\sqrt{2}} \right)^{1-b} - \frac{b-1}{b} \left(\frac{|f(\chi)|}{\sqrt{2}} \right)^{1-2b}  \right]
\end{equation}
\begin{equation}
    \psi \simeq \frac{\chi}{N\sqrt{2}\Gamma(1/b)} \mathrm{e}^{-\big( \frac{|f(\chi)|}{\sqrt{2}} \big)^b} \left[\left(\frac{|f(\chi)|}{\sqrt{2}} \right)^{2-b} - \frac{b-2}{b} \left(\frac{|f(\chi)|}{\sqrt{2}} \right)^{2-2b}  \right]
\end{equation}
giving, once plugged in Eq.~\eqref{equ:self_cons_appendix}
\begin{equation}
\begin{aligned}
    2\Gamma(1/b) \, \mathrm{e}^{\big( \frac{|f(\chi)|}{\sqrt{2}} \big)^b} =& \chi f(\chi) \left[\left(\frac{|f(\chi)|}{\sqrt{2}} \right)^{1-b} - \frac{b-1}{b} \left(\frac{|f(\chi)|}{\sqrt{2}} \right)^{1-2b}  \right]\\ 
    &+ \sqrt{2} \chi \left[\left(\frac{|f(\chi)|}{\sqrt{2}} \right)^{2-b} - \frac{b-2}{b} \left(\frac{|f(\chi)|}{\sqrt{2}} \right)^{2-2b}  \right].
\end{aligned}
\end{equation}
Finally, given the we know that $\beta^+ < 1$ in this regime, then we can simply realise that $f(\chi) = - |f(\chi)|$ and so terms on the right hand side cancel out to give
\begin{equation}
    \mathrm{e}^{\big( \frac{|f(\chi)|}{\sqrt{2}} \big)^b} = \frac{\chi}{\sqrt{2}b \Gamma(1/b)} \left(\frac{|f(\chi)|}{\sqrt{2}} \right)^{2-2b},
\end{equation}
the final equation to be approximated.

\section{\label{appendix:heterogeneous}Heterogeneous returns and growth rates}

To extend our model to more realistic conditions, it is important to generalise the calculations to variable values of $\mu_i$. In such a case, the long-only portfolio and population dynamics calculations are no longer strictly equivalent, as in the portfolio $\mu_i$ and $z_i$ are independent while regarding species we have $z_i = \mu_i/k_i$.

Starting with the long-only portfolio, we now have
\begin{equation}
    \theta_i = \Theta\left( \mu_i - \beta_i \frac{\sum \beta_j \mu_j/z_j}{\sum_j \beta^2_j/z_j + 1} \right).
\end{equation}
The first important aspect to notice here is that, just as was the case for $\beta$, we may fix the mean of $\mu$ to 1 without loss of generality as any multiplicative rescaling $\mu_i \to \alpha \mu_i$ leaves the above equation invariant. From there, we may proceed as before, now defining a threshold on the quantity $\phi = \beta/\mu$:
\begin{equation}
    \tilde{\phi}^+ = \frac{\sum_j \phi_j^2 \mu_j^2 \theta_j /z_j + 1}{\sum_j \phi_j \mu_j^2 \theta_j/z_j}.
\end{equation}
The probability for an asset to be included in the portfolio is now given by the function $F_N(\phi)$ that is qualitatively identical to the previously introduced equivalent for $\beta$ alone. Taking $\mu_i$ and $\beta_i$ to have the joint probability distribution $\rho(\beta,\mu)$, we must calculate
\begin{equation}
    \llangle \phi^k \mu^2 \rrangle = \int \dd \phi \int \dd \mu \int \dd \beta \, \phi^k F_N(\phi) \mu^2 \rho(\beta,\mu) \, \delta\left( \phi - \frac{\beta}{\mu} \right).
\end{equation}
From the results of \ref{appendix:sommerfeld}
, we know that this integral can be calculated as
\begin{equation}
    \llangle \phi^k \mu^2 \rrangle = \int_{-\infty}^{\phi^+} \dd \phi \, \phi^k h(\phi) + \mathcal{O}\left(\frac{1}{N^2} \right)
\end{equation}
where the key step is therefore calculating
\begin{equation}
    h(\phi) = \int \dd \beta \int \dd \mu \, \mu^2 \rho(\beta,\mu) \, \delta\left( \phi - \frac{\beta}{\mu} \right)
\end{equation}
that will act as an effective distribution of $\phi$. Assuming now that both $\beta$ and $\mu$ are distributed around 1 with a standard deviation scaling in $N^{-1}$, we may shift and rescale the problem by taking
\begin{equation}
    \beta = 1 + \frac{x}{N}, \qquad \mu = 1 + \frac{y}{N}, \qquad \phi = 1 + \frac{w}{N}
\end{equation}
and $\tilde{\rho}(x,y)$ that is the distribution of $\beta$ and $\mu$ centred now at 0 and with standard deviation of order 1. Using the scaling property of the Dirac delta distribution, we finally have the rescaled effective density for $\phi$ that is given by
\begin{equation}
\begin{aligned}
    \tilde{h}(u) &= \int \dd x \int \dd y \, \left(1 + \frac{2y}{N}\right) \tilde{\rho}(x,y) \, \delta(w-x+y) + \mathcal{O}\left( \frac{1}{N^2} \right) \\
    &= \int \dd y \, \tilde \rho(u+y,y) + \mathcal{O}\left( \frac{1}{N} \right),
\end{aligned}
\end{equation}
as it will be shortly apparent that the $N^{-1}$ contribution vanishes in the self-consistent equation. Note that when $\tilde \rho(x,y)$ is a bivariate Gaussian, $\tilde{h}(u)$ is also Gaussian.

Expressing the threshold in $\phi^+$ as
\begin{equation}
    \phi^+ = 1 + \frac{\tilde{f}}{N},
\end{equation}
where $\tilde{f}$ depends on the parameters describing the distribution $\tilde h$.
We may go back to the self-consistent equation that is analogous to that for $\beta^+$ and reads
\begin{equation}
    \phi^+ = \frac{\langle \phi^2 \rangle_c}{\langle \phi \rangle_c} + \frac{1}{N} \frac{1}{\langle \phi \rangle_c} + \mathcal{O}\left( \frac{1}{N^2} \right)
\end{equation}
with now
\begin{equation}
    \langle \phi^k \rangle_c = \int_{-\infty}^{\phi^+} \dd \phi \, \phi^k h(\phi) = \int_{-\infty}^{\tilde{f}} \dd u \, \left( 1 + \frac{ku}{N} \right) \tilde{h}(u) + \mathcal{O}\left( \frac{1}{N^2} \right).
\end{equation}
As such, we finally obtain an equation that is almost identical to the homogeneous $\mu_i = 1$ case, contributions of order 1 cancel out and we have, at order $N^{-1}$,
\begin{equation}
    \tilde{f} = \frac{1}{m} + \frac{1}{m} \int_{-\infty}^{\tilde{f}} \dd u \, u \tilde{h}(u)
\end{equation}
with
\begin{equation}
    m = \int_{-\infty}^{\tilde{f}} \dd u \, \tilde{h}(u).
\end{equation}
Clearly, the $N^{-1}$ term in the expression of $\tilde{h}(u)$ is dominated and therefore the $\mu^2$ term that was initially present turns out to be inconsequential.

With this result in mind, we can look at the equilibrium ecosystem problem. As previously mentioned, the relation $z_i = \mu_i/k_i$ means that the threshold is now given by
\begin{equation}
    \tilde{\phi}^+ = \frac{\sum_j \phi_j^2 k_j \mu_j \theta_j + 1}{\sum_j \phi_j k_j \mu_j \theta_j}.
\end{equation}
Thus, leaving aside the $k_j$ that play no part in the inclusion or not of the species (as was the case for $z_j$), the problem is identical to the long-only portfolio problem, albeit with $\llangle \phi^k \mu \rrangle$ to be calculated instead of $\llangle \phi^k \mu^2 \rrangle$. Having just determined that when taking $\mu$ to be distributed in an $N^{-1}$ region about its mean its contribution in the integral is negligible, we find ourselves with exactly the same self-consistent equation as above.

\section*{References}
\bibliographystyle{unsrt}
\bibliography{bibs}

\end{document}